\documentclass[preprint]{revtex4-2}
\usepackage{amsmath,amssymb}
\usepackage{graphicx}
\usepackage{physics}
\usepackage{natbib}
\usepackage{amsfonts}
\usepackage{lineno}
\usepackage[version=3]{mhchem}

\begin{document}
    \title{Onset of slow dynamics in dense suspensions of active colloids}

\author{Antina Ghosh$^{1}$, Sayan Maity$^1$, Vijayakumar Chikkadi$^{1}$}

\affiliation{
$^{1}$ Indian Institute of Science Education and Research Pune, Pune 411008, India\\
}

\begin{abstract}
Slow relaxation and heterogeneous dynamics are characteristic features of glasses. The presence of glassy dynamics in nonequilibrium systems, such as active matter, is of significant interest due to its implications for living systems and material science. In this study, we use dense suspensions of self-propelled Janus particles moving on a substrate to investigate the onset of slow dynamics. Our findings show that dense active suspensions exhibit several hallmark features of slow dynamics similar to systems approaching equilibrium. The relaxation time fits well with the Vogel-Fulcher-Tamman (VFT) equation, and the system displays heterogeneous dynamics. Furthermore, increasing the activity leads to faster relaxation of the system, and the glass transition density predicted by the VFT equation shifts to higher densities. The measurement of the cage length and persistence length reveal they are of the same order over the range of activities explored in our study. These results are in agreement with recent particle simulations.
\end{abstract}


\maketitle

\section{Introduction}
When molecular liquids are cooled, their timescales of structural relaxation increases. If we avoid crystallization by employing rapid cooling rates, most liquids fall out of equilibrium to enter a metastable phase of supercooled liquids. Subsequently, they form glasses at temperatures lower than the glass transition temperature $T_g$ \cite{Stillinger01, Biroli11}. A similar phenomenon occurs in Brownian colloidal suspensions, where the colloidal glass transition is controlled by the particle density \cite{Pusey91, Weeks12, Weitz00, Megen86}. As density increases, particle motion becomes increasingly hindered by the cages formed by neighboring particles, leading to a slowdown in dynamics. In recent years, the emergence of slow dynamics in active matter has garnered considerable attention due to its relevance to living systems \cite{Simha13}. For instance, the collective motion of cells in tissues during embryo development \cite{Manning13}, wound healing \cite{Trepat11}, and cancer metastasis exhibits glass-like features such as slow relaxation and heterogeneous dynamics \cite{Kas17, Kas21}. Bacterial cell cytoplasm shares several properties with glass-forming liquids \cite{Jacobs-Wagner14}. A central question in these investigations is whether systems driven far from equilibrium display slow glassy dynamics and how does it compare to systems close to equilibrium \cite{Kurchan13}. In this article, we shed new light on some aspects of slow dynamics in dense active matter by performing experiments using a monolayer of active colloids.

Self-propelled particles have served as model systems for understanding various aspects of dense active matter \cite{Bechinger12,Janssen19}. They have been studied extensively in simulations using two well-established models of active Brownian particles (ABPs) \cite{Marchetti12, Schimansky-Geier12} and active Ornstein-Uhlenbeck particles (AOUPs) \cite{Szamel14, Leonardo15}. In these models the particles undergo overdamped motion in a viscous fluid, and their self-propulsion is described using a force that remains either constant or evolves over time. In the dilute limit, both models depict a persistent random walker, with the mean-square displacement (MSD) given by: $\left< \Delta r^2 (t)\right> = 6Tt + 6T_a\left[\tau_p(e^{-t/\tau_p}-1+t)\right]$, where $T$ represents the thermodynamic temperature, $T_a$ is the active temperature, and $\tau_p$ is the persistence time of the random walker \cite{Golestanian07}. The active temperature $T_p=f^2\tau_p/3$ for ABPs. In the limit of short times when $t << \tau_p$, the MSD $\left< \Delta r^2 (t)\right> \approx 6Tt +3T_at^2/\tau_p$, indicating a quadratic dependence on $t$ due to ballistic motion. However, in the limit of long times $t >> \tau_p$, the MSD $\left< \Delta r^2 (t)\right> \approx 6(T_a+T)t \equiv 6T_{eff}t$, which increases linearly with $t$, highlighting diffusive motion. These dilute-limit predictions were first confirmed by Howse and coworkers \cite{Golestanian07} and have since been validated by other experiments using  colloidal models \cite{Bocquet12,Bechinger12,Volpe16}.

Investigations into active matter systems at high densities and the onset of slow dynamics have primarily focused on theoretical models and numerical simulations. The earliest studies \cite{Dijkstra13, Kurchan13, Berthier14} revealed that slow dynamics in active systems exhibit all the essential features of supercooled liquids approaching an equilibrium glass transition. These features include caging, dynamical slowing down, non-exponential time correlation functions, and dynamic heterogeneity.
Particle simulations of active systems based on hard-sphere models have demonstrated a monotonic effect of activity, showing that an increase in activity pushes the glass transition to higher densities \cite{Dijkstra13, Berthier14}. However, subsequent simulations using Lennard-Jones potentials, which include an attractive component, revealed a non-monotonic effect of activity. In these simulations, activity could either enhance or suppress the fluidization of the system \cite{Berthier15, Dasgupta16, Berthier16}.
Further simulations conducted by Berthier and coworkers \cite{Szamel17}, using a model that allowed continuous interpolation from hard-sphere-like repulsive interactions to Lennard-Jones-type models with an attractive component, clarified the contradictory effects of activity. In the hard-sphere limit at low temperatures and moderate densities, activity had a monotonic impact on the relaxation time of the system, leading to faster relaxation with increased activity. However, at moderate temperatures and high densities (resembling Lennard-Jones models), activity exhibited a non-monotonic impact on relaxation. This study showed that interparticle interactions and the structure changes induced by activity have a significant influence on the onset of slow glassy dynamics. 
A recent study\cite{Janssen21} identified that the long-time diffusion constant increases with increasing persistence time when it is small compared to the cage length. An opposite effect is observed when the cage length is smaller than the persistence length. Active glasses are reported to display other interesting features, including multiple aging regimes for highly persistent forces \cite{Sollich20, Janssen22}, and tunable fragility due to doping of active particles in passive glasses \cite{Dasgupta16}.

Experimental investigations of this topic are scarce. The earliest experimental reports of glassy dynamics in active matter involved living systems consisting of tissues \cite{Manning13, Trepat11, Kas17, Kas21}. As far as synthetic systems are concerned, a recent study \cite{Ganapathy22} reported the role of topological defects and the re-entrant effect of activity in dense active matter consisting of granular ellipsoidal particles. The only experimental study with colloidal spheres was conducted by Klongvessa and co-workers \cite{Leocmach19-1, Leocmach19-2}, where they reported a non-monotonic influence of activity on the dynamics of the system.
Their primary findings were that increasing activity leads to faster relaxation in an ergodic state at lower densities. However, in a non-ergodic state at higher densities, activity shows a non-monotonic effect on relaxation time. Initially, relaxation time increases with higher activity levels, but decreases again at higher activity intensities. The platinum-coated gold colloids used in their study exhibit characteristics similar to soft particles, such as poly(N-isopropyl acrylamide) (PNiPAM) microgel particles \cite{Cipelletti18}. It is important to clarify whether these conclusions apply generically to quasi-hard particle systems. Recent simulations \cite{Janssen21, Janssen22-2} suggest that the relative sizes of the cage and the persistence length of active particles critically influence the activity's impact. However, these numerical results lack experimental validation. Furthermore, the nature of slow dynamics in quasi-hard particle systems remains to be explored.

In this article, we present an experimental investigation of the onset of slow relaxation in a dense layer of photoactive colloids. The strength of activity is varied by tuning the intensity of UV light. In the absence of activity, the relaxation time of the system increases monotonically, suggesting quasi-hard particle nature of our system. Our experiments reveal a monotonic effect of activity on the relaxation time over the range of densities and activities studied in our experiments. Increasing the activity leads to faster relaxation. The relaxation time over a range of densities and at various magnitudes of activity is well described by a Vogel-Fulcher-Tamann (VFT) relation. The glass transition densities obtained from the VFT fits reveal that increasing the activity shifts the glass transition density to higher values. The examination of structural changes and length-scale of cooperative motion due to activity point to faster relaxation due to activity. The investigation of the ratio of cage size and persistence length due to activity reveals that our results are in agreement with simulation results \cite{Dijkstra13,Janssen21,Szamel17}. 

\section{Experimental systems}
The active colloids in our experiments are light-driven, self-propelled Janus particles composed of \ce{SiO_2} and anatase \ce{TiO_2} halves \cite{Singh17}. The details of the experimental methods followed is given in the supplementary information. To obtain amorphous configurations, we have used a binary mixture of particles with diameters of $\sigma_s=2.64~\mu m$ and $\sigma_l=3.12~\mu m$. The ratio of small to large particles in our system is in the range 2:1, and all the analysis was performed using small particles. The Fig.1a shows an image of the dense monolayer of the particles that was obtained by sedimenting particles in a circular cavity of $30 \mu m$ thickness, fabricated using photolithography techniques. The diameter of the circular cavity is $150 \mu m$. An array of such cavities was created to study the systems over a range of densities. The particles only in the central region or in the unshaded region were considered for analysis to prevent the influence of walls. When the Janus particles are dispersed in an aqueous hydrogen peroxide solution (\ce{H_2 O_2} $3\%$, \ce{pH}$\sim7$), they display passive Brownian diffusion in the absence of UV light. However, upon illumination of UV light of wavelength $365~nm$ the Janus colloids show self-propulsion giving rise to ballistic motion on short time scales and a diffusive motion on long  time scales. The propulsion arises from the photocatalytic decomposition of \ce{H_2O_2} at the \ce{TiO_2} surface by UV-promoted electron–hole pairs \cite{Sen10, Singh17}. Since the rate of photocatalytic decomposition near the \ce{TiO_2} surface depends on the intensity of light, the velocity of Janus particles increases with UV power. The details about these aspects are provided in the supplementary information. The rotational diffusion time of the particles is $\tau_R\sim3s$, which was measured from the MSD of particles (see supplementary information).
The UV power is varied in our experiments to obtain velocities in the range of $0.05 \mu m/s $ to $0.2 \mu m/s$. The strength of activity in our study is expressed using an effective temperature, $T_{eff}$, see supplementary information for details. The density of particles is represented by the area fraction $\phi \sim N \pi \sigma^2/4A$, where A is the area of the field of view.

\begin{figure}[h!]
\centering
\includegraphics[width=.24\textwidth]{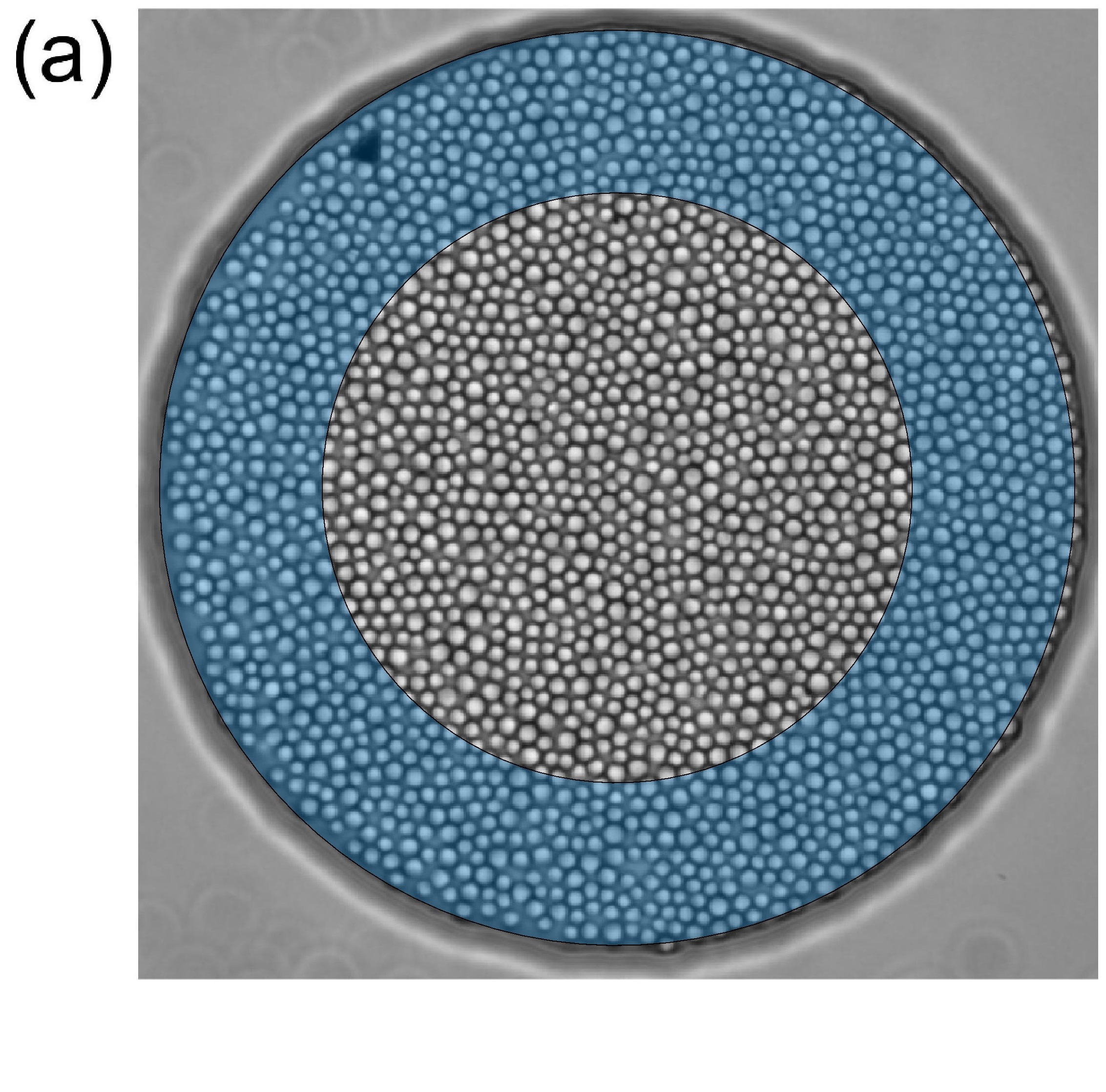}
\hspace{2mm}
\includegraphics[width=.34\textwidth]{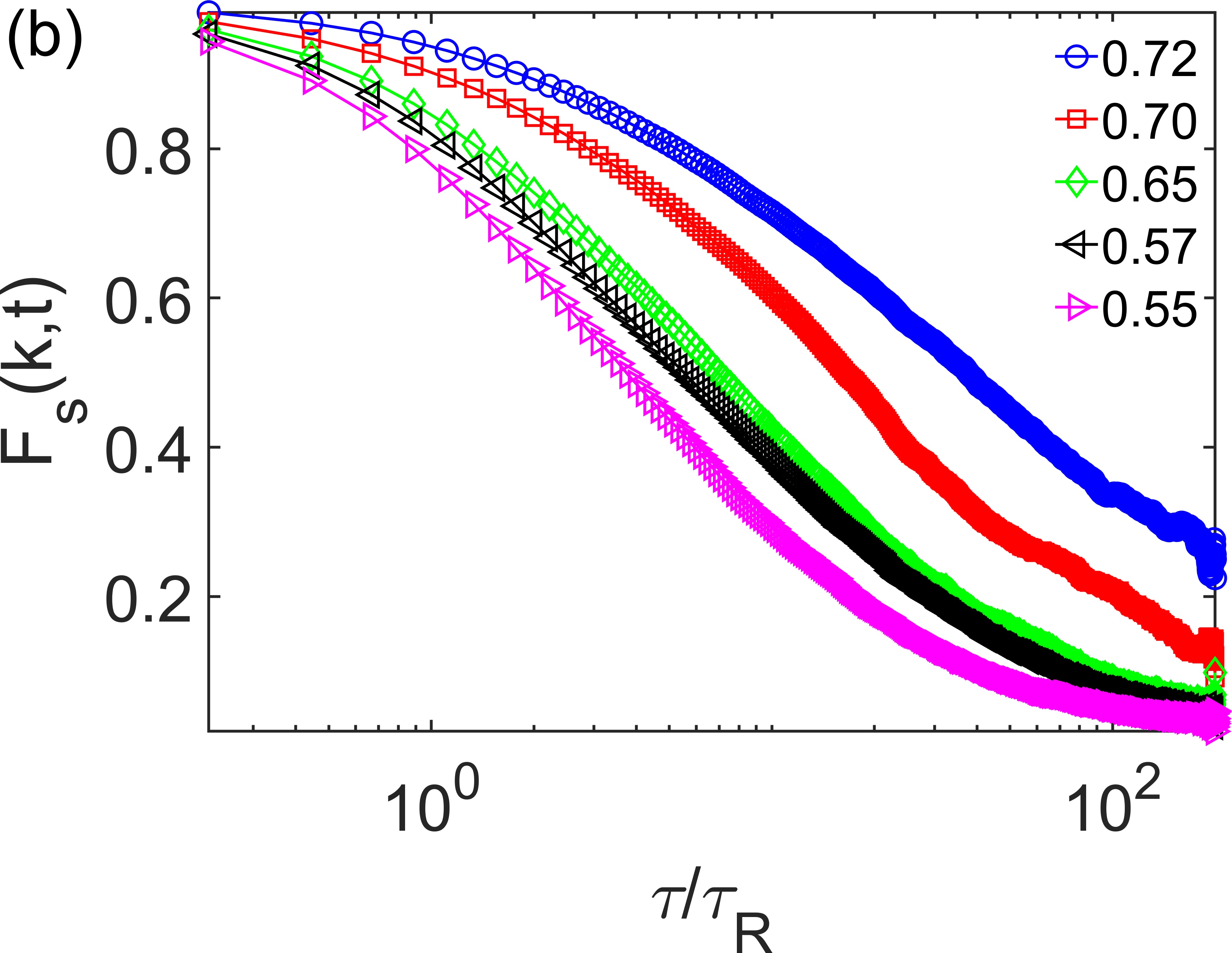}
\hspace{2mm}
\includegraphics[width=.27\textwidth]{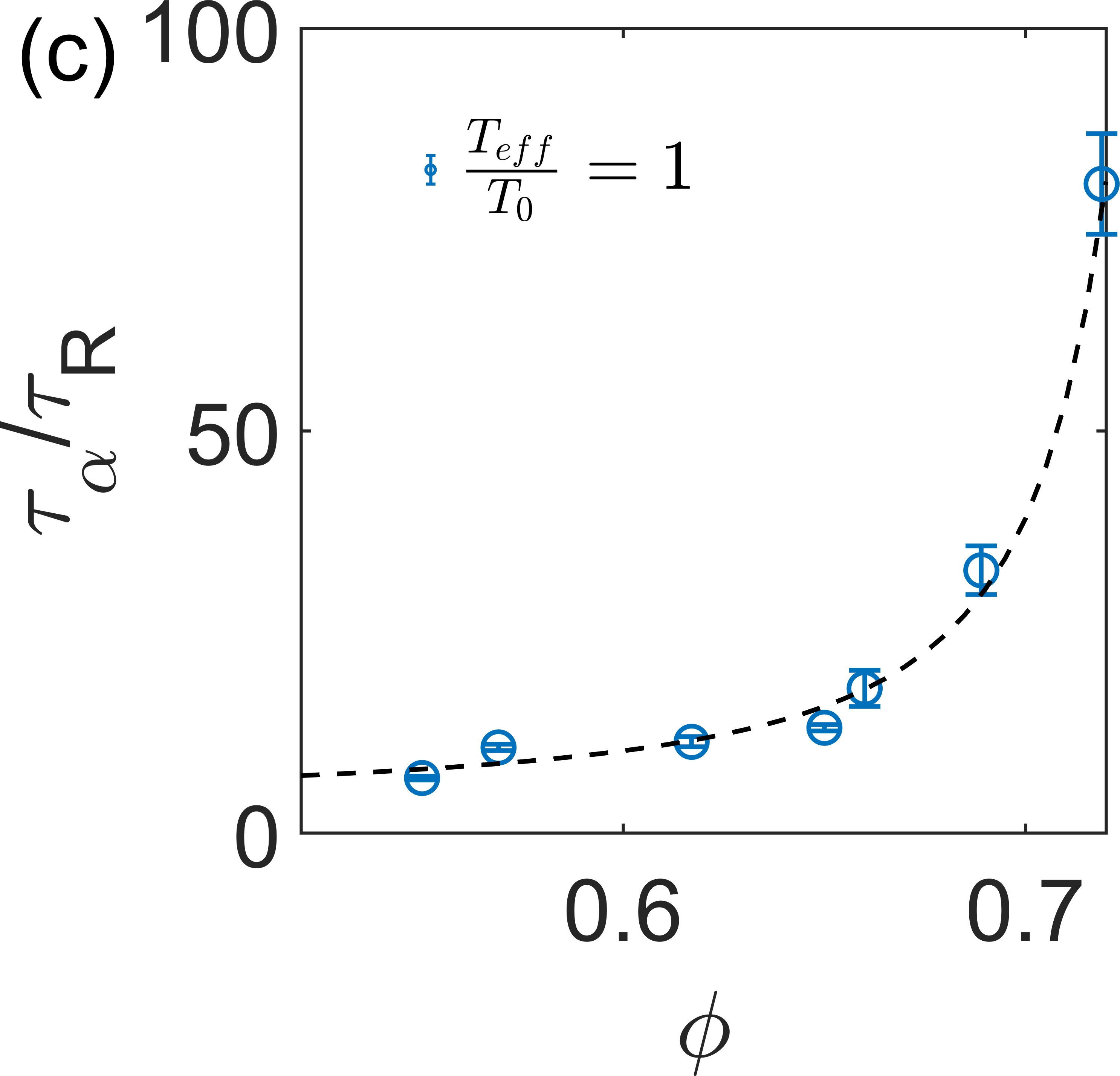}
\caption{Relaxation of passive Brownian system in the absence of activity, corresponding to $\frac{T_{eff}}{T_0}=1$. (a) A bright-field image of circular cavity with binary colloids. The diameter of the cavity is $\sim 150 \mu m$ and it's depth is $\sim 30 \mu m$. The particles in the unshaded region of diameter $90\mu m$ are considered for analysis. (b) Self part of the intermediate scattering function $F_s(k,t)$, where $k=2\pi/\sigma_s$, over a range of area fractions of the colloidal particles from $\phi\sim0.55-0.72$. The x-axis is scaled by the rotational diffusion time scale $\tau_R$. (c) The relaxation time $\tau_{\alpha}$ is plotted as a function of the area fraction of colloidal particles. The thick line is a fit to the data of the form .}
\label{Fig1}
\end{figure}

\section{Onset of slow dynamics in the absence of activity}
When the density of system is small, the particles diffuse freely and the relaxation is fast. With increasing density, the particles motion is hindered due to the cage formed by nearest neighbors, leading to the onset of slow dynamics in the system. The dynamics of particles within the cage and the their escape from it is well captured by the self part of the intermediate scattering function, $F_s(k,t)$, given by 
\begin{equation}
F_s(k,t)=\left<\frac{1}{N}\sum_{i=1}^{N} \textnormal{exp} \left[ j\textbf{k} \cdot \Delta \mathbf{r}_i(\tau) \right] \right>,
\end{equation}
where the $k=2\pi/\sigma_l$ and $\Delta\mathbf{r}(\tau)$ is the cage-relative displacement \cite{Weeks17} defined with respect to $N_i$ nearest neighbors in the following way: 
\begin{equation}
\Delta \mathbf{r}_i(\tau) = \mathbf{r}_i(t + \tau) - \mathbf{r}_i(t) - \frac{1}{N_i}\sum_i^{N_i}[\mathbf{r}_j(t + \tau) - \mathbf{r}_j(t)].
\end{equation}
Previous studies \cite{Weeks17, Keim17} have demonstrated long-wavelength fluctuations in two-dimensional disordered colloidal systems leading to large particle displacements without neighbor changes. Investigating the relative displacement of particles relative to their first neighbors provided a consistent understanding of the relaxation dynamics and the glass transition in both two and three dimensions. Therefore, the displacements in our study are computed relative to their first nearest neighbors, which we refer to as relative displacements in the rest of the article. The main panel of Fig.1b presents the $F_s(k,t)$ computed using relative displacements over a range of area-fractions from $\phi\sim0.55-0.72$ in the absence of activity, which is due to Brownian motion. The slowing down of the dynamics is apparent from the slower decay of $F_s(k,t)$ with increasing density. The relaxation time scale $\tau_{\alpha}$ is estimated from the time taken for $F_s(k,t)$ to decay by a factor of $1/e$. It is shown as a function of the area fraction of colloids in Fig.1c. The thick line in the figure shows a Vogel-Fulcher-Tamann fit of the form 
$\tau_{\alpha} = \tau_R~\textnormal{exp} \left[ \frac{A}{(\phi_c-\phi)} \right]$, $\phi_c$ is the critical area-fraction where $\tau_{\alpha}$ diverges. The best-fit procedure yields a critical density of $\phi_c=0.79\pm0.012$, which reasonably agrees with earlier simulation studies \cite{Weeks09}. The non-exponential relaxation of density correlation is apparent from the form of $F_s(k,t)$, and it arises from the multiple relaxation time scales of the system. To confirm the onset of slow glassy dynamics in our system, we have tested for the signatures of aging at $\phi\sim0.72$ and $0.55$, the two extreme densities in our study. The mean square displacement (MSD) of particles was plotted for different waiting times. The dynamics were found to be dependent on the waiting time at $\phi\sim0.72$, while at $\phi\sim0.55$, the MSD curves overlapped on each other. See supplementary information for details. These results establish the onset of glassy dynamics at higher densities in our system. 

\begin{figure}[h!]
\centering
\includegraphics[width=.37\textwidth]{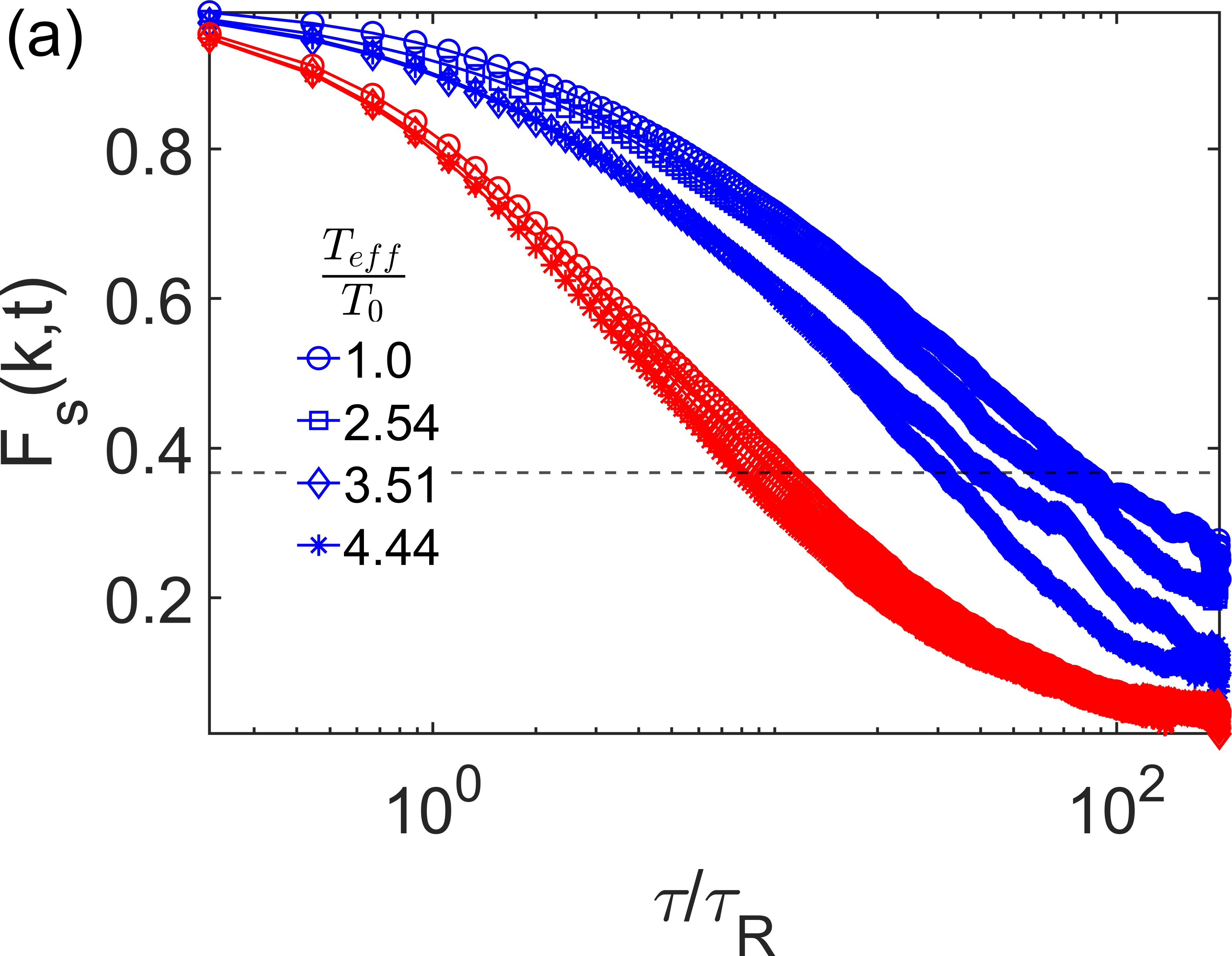}
\includegraphics[width=.29\textwidth]{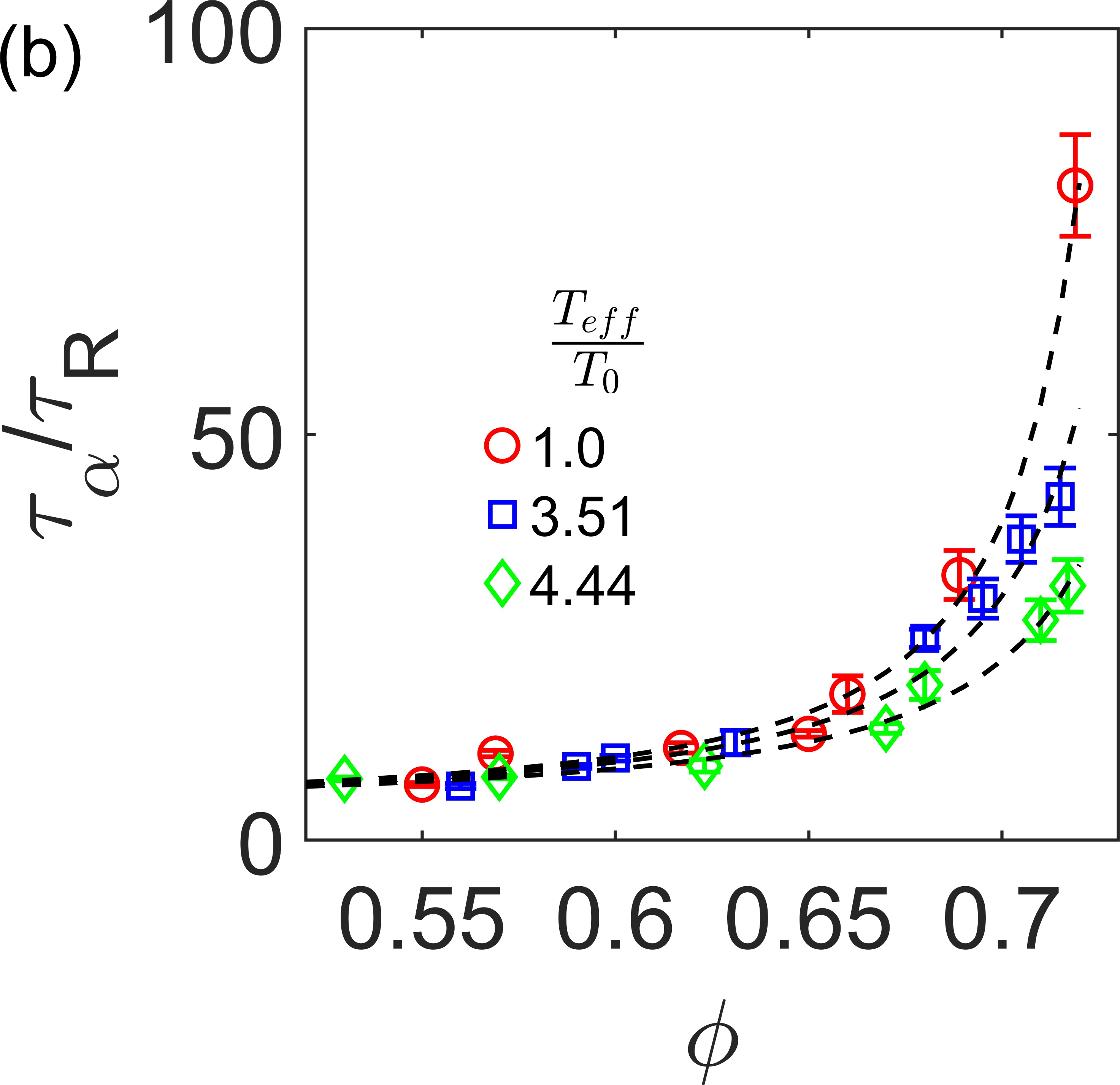}
\includegraphics[width=.29\textwidth]{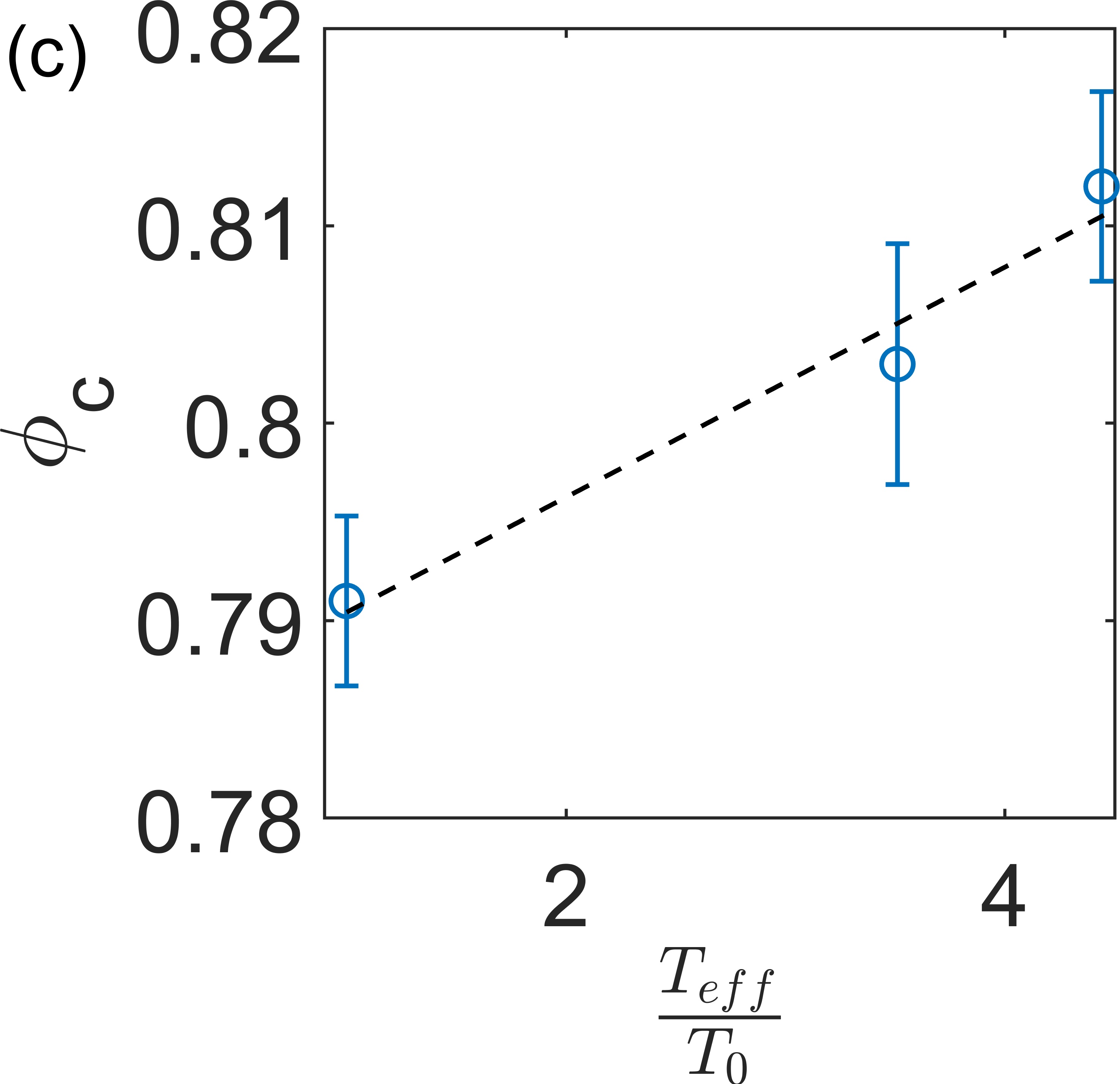}
\caption{Relaxation of active systems. (a)The self part of the intermediate scattering function $F_s(k,t)$ at various effective temperatures ranging from $\sim1-4.44$, and at two different area fractions $\phi\sim0.72$ and $\phi\sim0.55$. Different marker symbols are used to distinguish the effective temperatures, and the colors distinguish area fractions. (b) The relaxation time $\tau_{\alpha}$ of the system is presented as a function of area fraction of colloids at various effective temperatures. The marker symbols indicate experimental data and the thick lines are the VFT fit to the curves. (c) The critical area fraction $\phi_c$ obtained from the fits in (b) is shown as a function of the effective temperature $T_{eff}$. } 
\label{Fig2}
\end{figure}

\section{Influence of activity on the dynamics}
We present the influence of activity on the dynamics of the system in this section. As discussed earlier, the activity of our systems is quantified using an effective temperature and it is controlled by the intensity of UV light. The effective temperature is varied from $T_{eff}\sim1-4.4$, and the intermediate scattering function $F_s(k,t)$ show in Fig.2a at $\phi\sim0.72$ and $0.55$ reveals a monotonic decrease of the relaxation time $\tau_{\alpha}$ with increasing activity. What is also apparent is that the activity has a stronger effect at $\phi\sim0.72$ compared to $\phi\sim0.53$. The relaxation time $\tau_{\alpha}$ decresases by a factor of $3$ at $\phi\sim0.72$. We next present the effect of activity over a range of area fractions for various effective temperatures in Fig. 2b. The lines show a VFT fit to the curves. The fit provides an estimate of the critical density $\phi_c$ where $\tau_{\alpha}$ diverges. We plot $\phi_c$ as a function of the effective temperature $T_{eff}$ in Fig.2c. Clearly, the critical area-fraction $\phi_c$ is pushed to higher densities with increasing activity over the entire range of the area fractions and activities investigated in our experiments. These results are in agreement with simulations of active particles performed using hard-sphere like interactions \cite{Dijkstra13}. Unlike earlier experiments \cite{Leocmach19-1,Leocmach19-2} using Janus colloids as active particles, which showed a non-monotonic variation of relaxation times $\tau_{\alpha}$ due to activity in the nonergodic states, our experiments reveal a monotonic variations over the range of effective temperatures explored in our experiments. These results highlight the fact that interparticle interactions play a significant role when the effect of activity is considered on the onset of slow dynamics.

\begin{figure}[!]
\centering
\includegraphics[width=.35\textwidth]{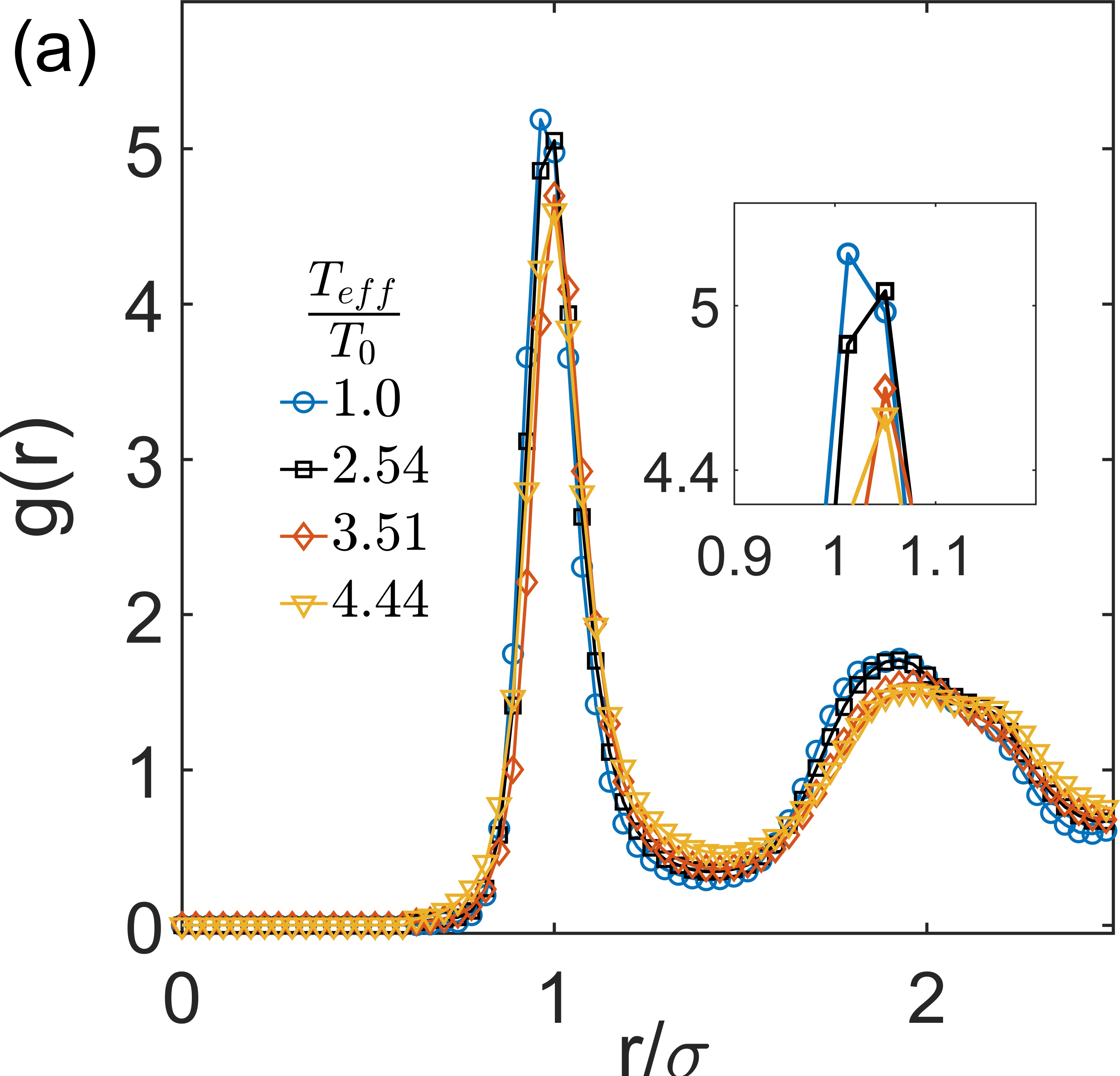}
\hspace{2mm}
\includegraphics[width=.33\textwidth]{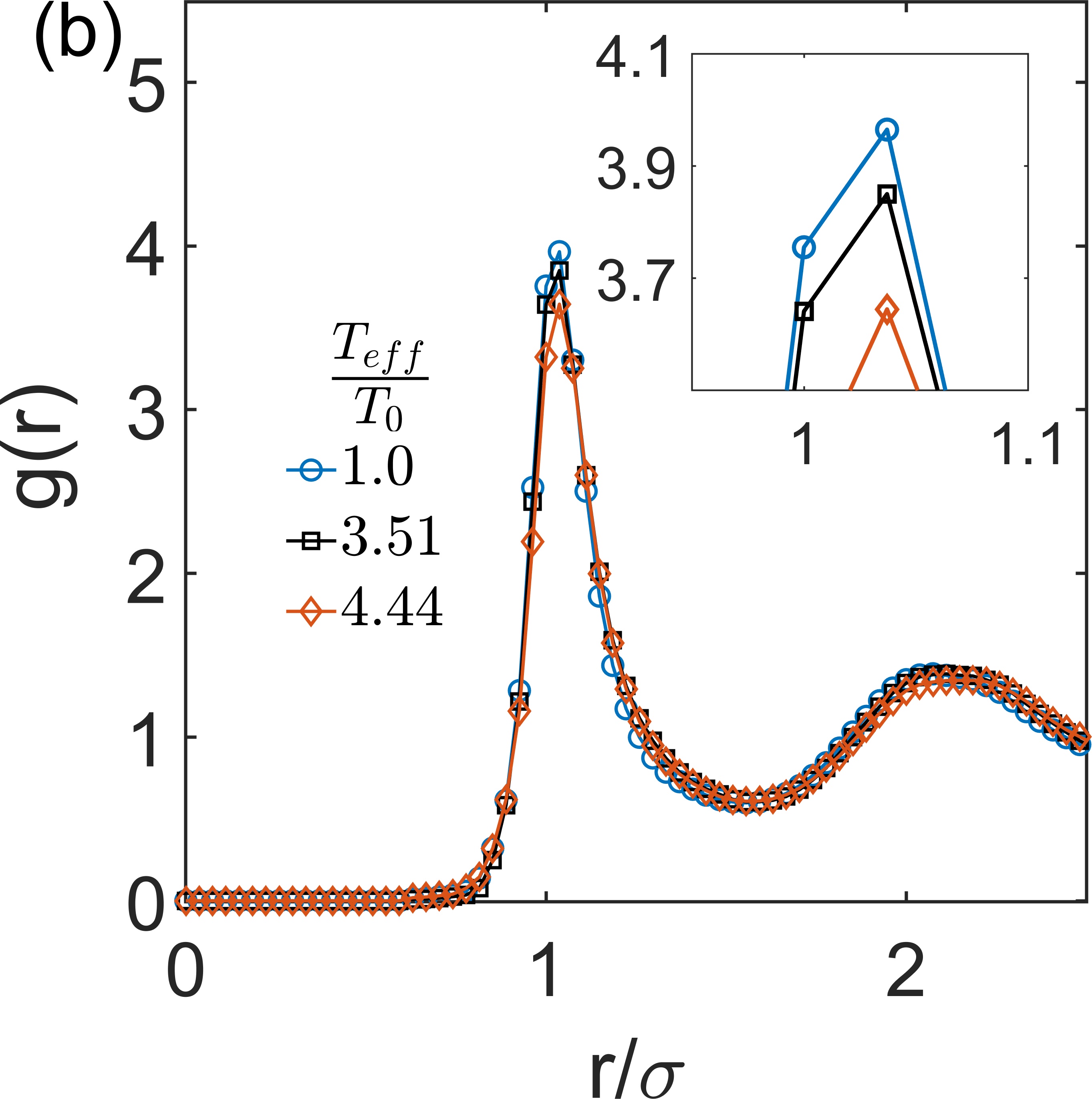}
\caption{Structural changes due to activity. The pair correlation function $g(r)$ of the system at different effective temperatures is presented at $\phi\sim0.72$ (a) and $\phi\sim0.55$ (b). The insets show a magnified view of the first peak.}
\label{Fig2}
\end{figure}

\section{Structural changes due to activity}
\subsection{Pair correlation function }

In this section, we investigate the effect of activity on the structure. We elucidate the effect of activity by computing the pair correlation function of small particles in the system,
\begin{equation}
g_{\alpha\beta}(r)=\frac{A}{N_\alpha N_\beta}\left<\sum_i^{N_\alpha}\sum_{j\ne i}^{N_\beta}\delta(r-|\mathbf{r}_i-\mathbf{r}_j|)\right>\\
\end{equation}
where the $N_{\alpha}$ and $N_{\beta}$ is the number of particles of different species. Only small particles were considered for all the calculations, so we denote $g_{ss}(r)$ simply as $g(r)$. The Fig.3a presents $g(r)$ for different effective temperatures at $\phi\sim0.72$. The effect is apparent from the variations in the height and position of the peaks. A magnified view of the first peak in the inset shows a decrease in the height of the peak with increasing activity. Even though the changes are small, there is a systematic variation due to activity. Similar changes are observed in the second peak too. All these observations point to restructuring of particles to achieve higher diffusivity in the system. The main panel in Fig.3b presents the changes in $g(r)$ due to activity at a lower area fraction of $\phi\sim0.55$. The corresponding inset presents a magnified view of the first peak. The variations in peak height appear to be smaller. A comparison of the changes in the Figs.3a and 3b suggests that the effect of activity on the structure is pronounced at higher densities.

\begin{figure}[h!]
\centering
\includegraphics[width=.26\textwidth]{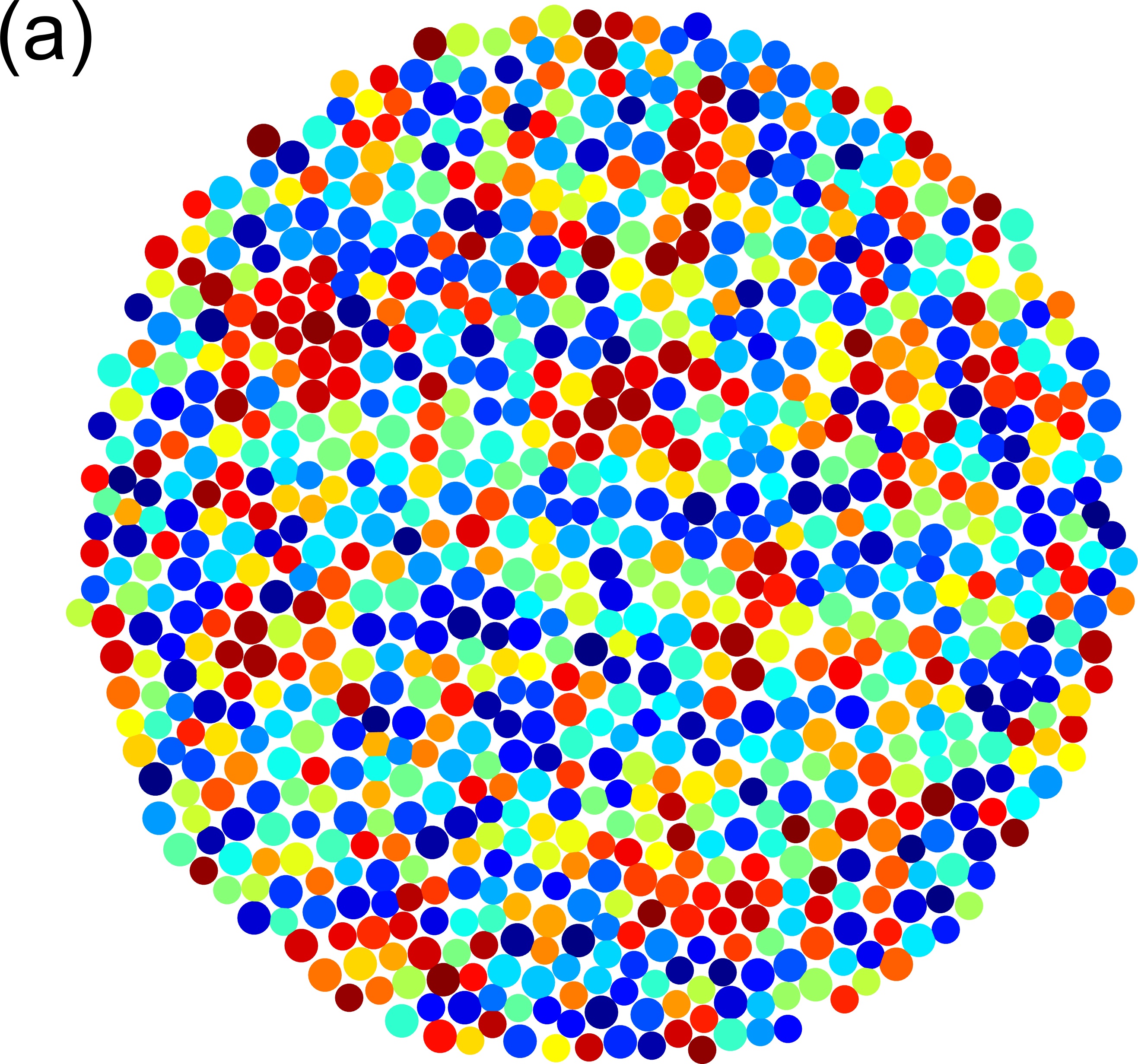}
\includegraphics[width=.31\textwidth]{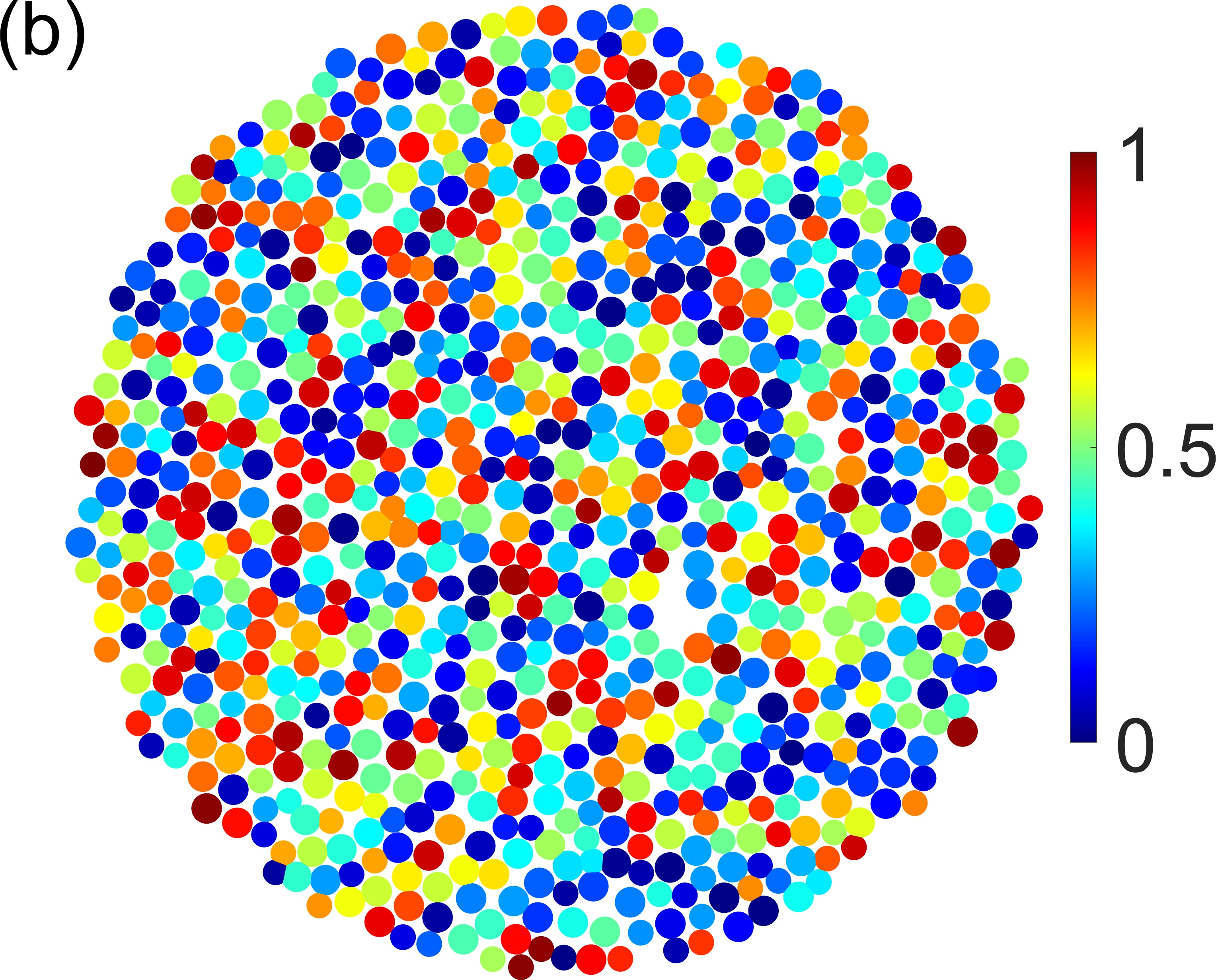}
\includegraphics[width=.34\textwidth]{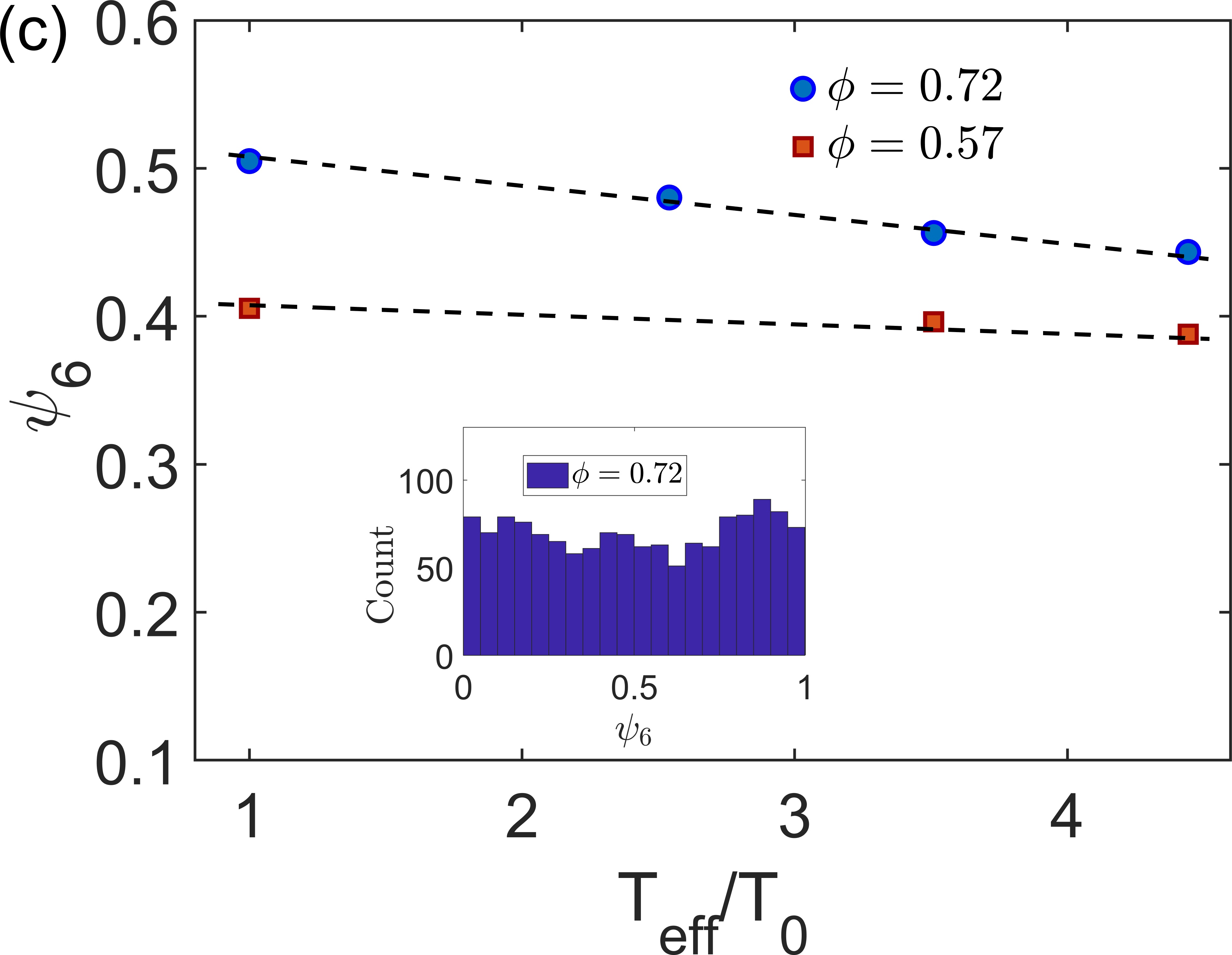}
\caption{Effect of activity on the local hexagonal ordering. (a) \& (b) Snapshots of the local hexagonal order of particles at an area fraction of $\phi\sim0.72$ and varying effective temperatures. The particles are color coded based on their value of $\psi_6$, and the panels in (a) and (b) are at effective temperatures $T_{eff}=1.0, \textnormal{ and } 4.44$, respectively. (c) The main panel shows the average $\psi_6$ as a function of $T_{eff}$ at area fractions $\phi\sim0.72$ and $0.55$. Different symbols are used to distinguish the area fractions. The inset is the histogram of $\psi_6$ at $\phi\sim0.72$  and $T_{eff}=1$.}
\label{Fig3}
\end{figure}

\subsection{Local ordering of particles}
We continue our investigation of the effect of activity on the structure by studying the local ordering of particles. In particular, we quantify the hexagonal order using local bond-orientational order parameter \cite{Halperin79} defined as,
\begin{equation}
\psi_{6}^{i}=\frac{1}{6}\sum_{j=1}^{N_f}\textnormal{exp}(i6\theta_{ij}),
\end{equation}
where $N_f$ is number of first nearest neighbors and $\theta_{ij}$ is the angle made by the bond between particles $i$ and $j$ with the $x-$axis. The local $\psi_6$ calculation were done using all kinds of particles. A color coded representation of local ordering is shown in Fig.4a and Fig.4b, where the particles are colored based on their $\psi_6$ values at $\phi\sim0.72$ and $T_{eff}=1.0$ and $4.44$, respectively. Red color indicates crystal-like hexagonal order and blue color indicates its absence. The presence of ordered patches are apparent and it appears to diminish with increasing activity in Fig.4a and Fig.4b. For a better understanding, a distribution of $\psi_6$ is shown in Fig.4c at $\phi $ and $T_{eff}$. Note that symbols are used to distinguish the activity and colors identify the area-fraction $\phi_s$. The peak of distribution $P(\psi_6)$ shifts to smaller values of $\psi_6$ with increasing activity. This effect is also clear from the inset of Fig.4c, which shows the mean of the distributions as a function of activity ($T_{eff}$). 

\begin{figure}[t]
\centering
\includegraphics[width=.24\textwidth]{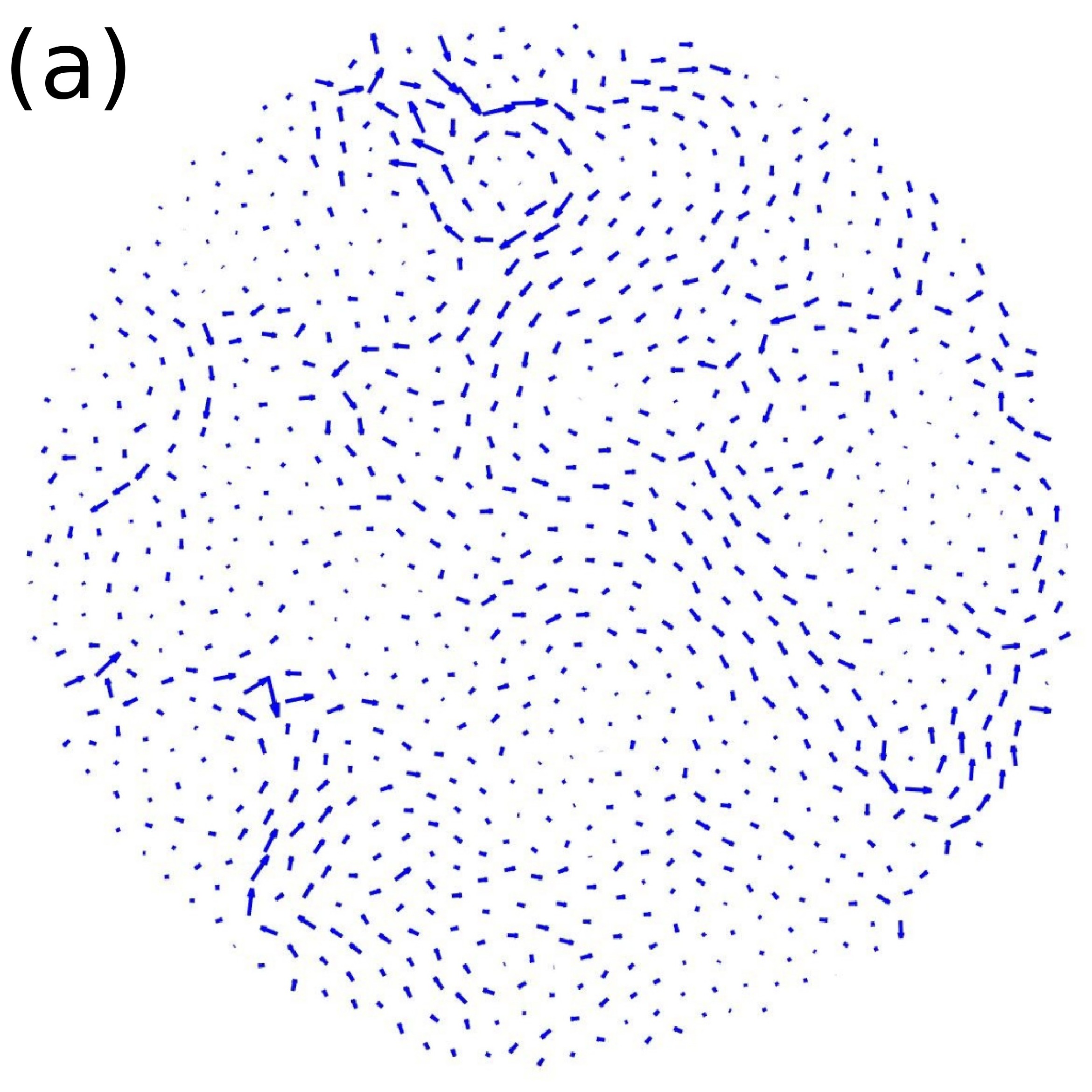}
\includegraphics[width=.24\textwidth]{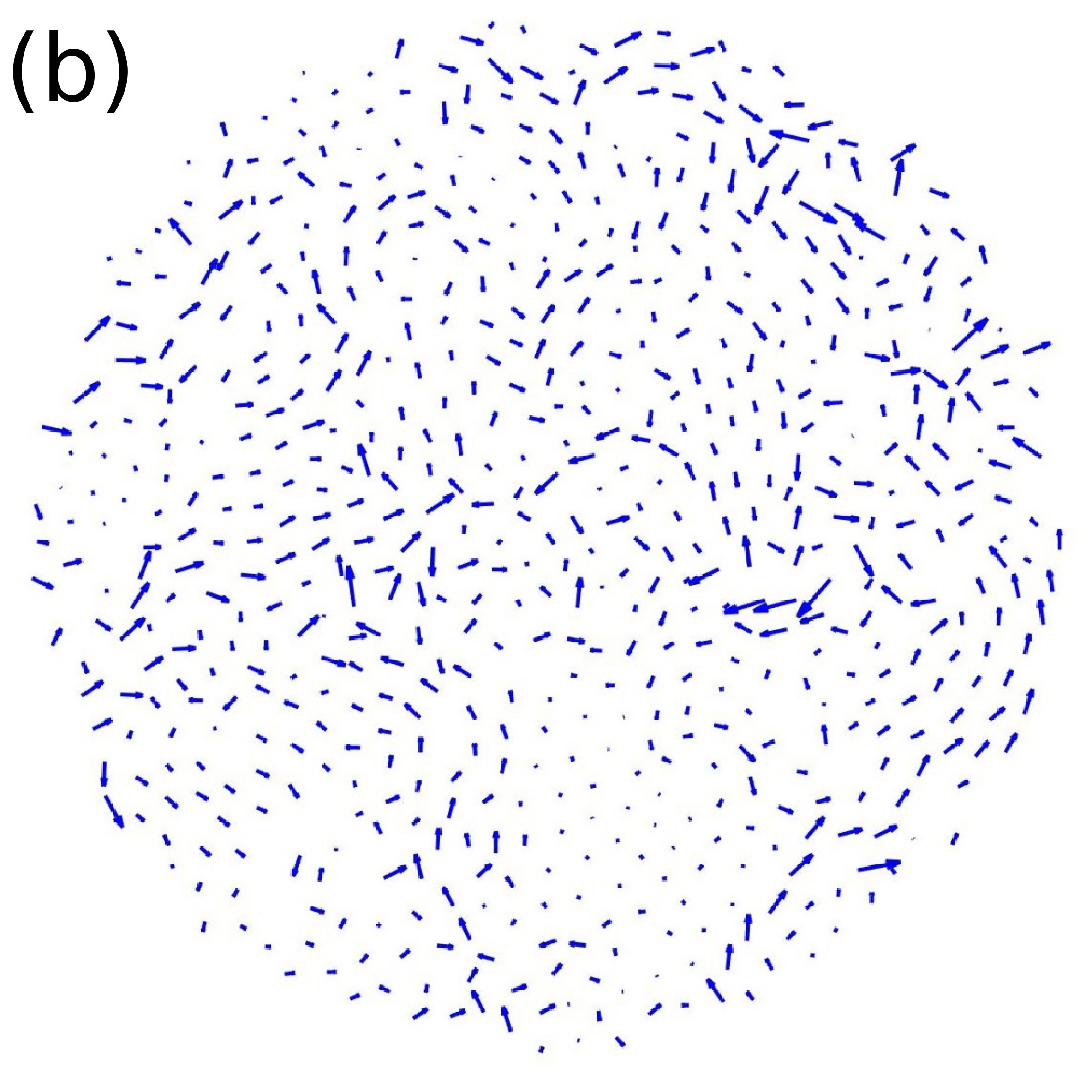}
\includegraphics[width=.37\textwidth]{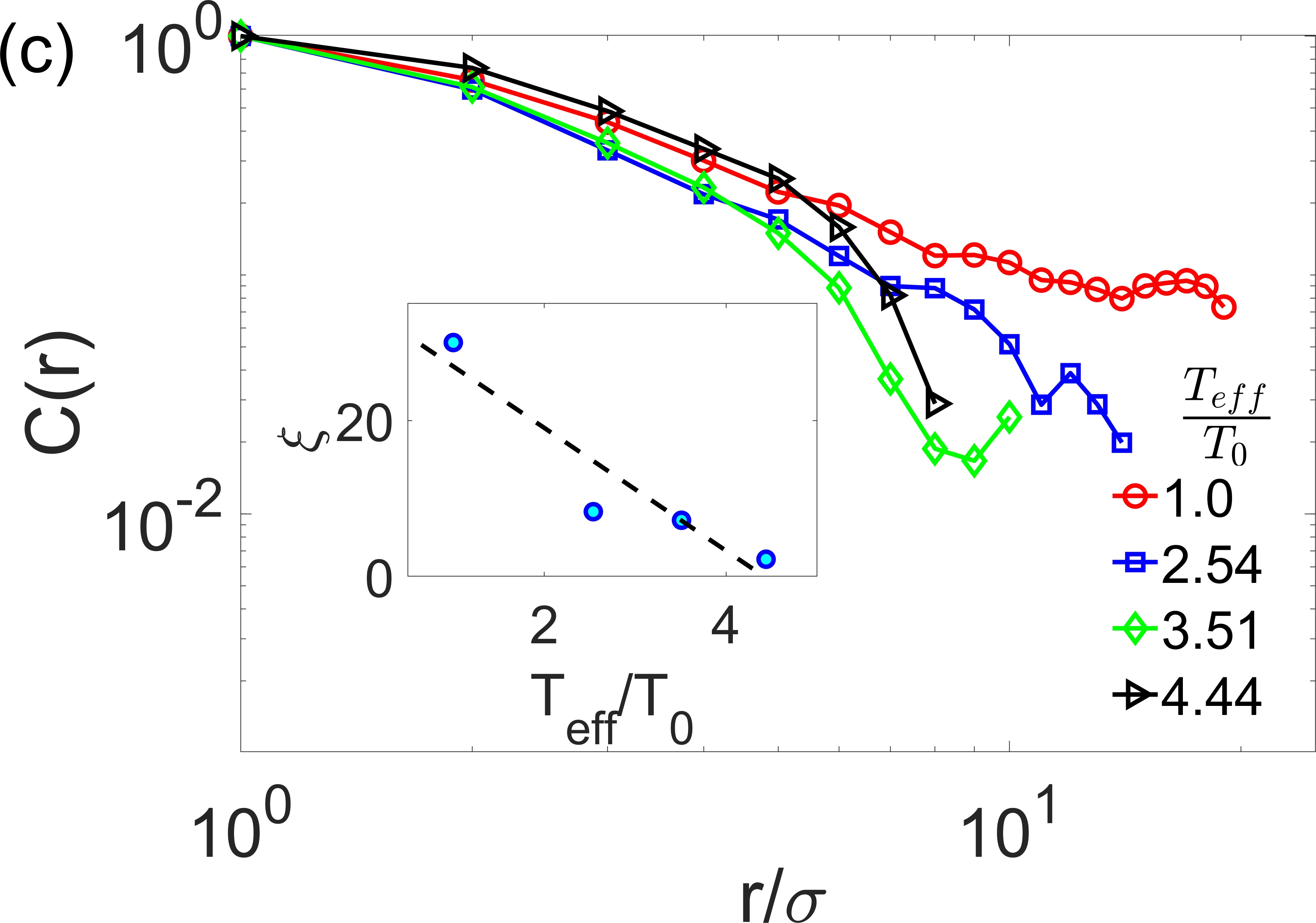}
\caption{Displacement fields and correlations that characterize spatially heterogeneous dynamics in dense systems of active colloids. The displacement vectors of particles are computed over a time scale $\tau_{\alpha}$ for $\phi\sim0.72$ at $T_{eff}=1$ (a) and $T_{eff}=4.44$ (b). (c) The displacement correlations for $\phi\sim0.72$ are shown at various effective temperatures $T_{eff}$ in the main panel. The inset shows the correlation lengths $\xi$ extracted from the correlation functions in the main panel.}
\label{Fig5}
\end{figure}

\section{Spatially heterogeneous dynamics}
One of the hallmark features of passive super-cooled liquids and passive glasses is the spatially heterogeneous dynamics \cite{Ediger00}. This means the mobility of particles differs significantly from one region to another region of the system, leading to clusters of slow and fast moving particles. The increase in the relaxation time of system with increasing density is related is attributed to the growth of length-scales of cooperative motion in the system \cite{}. We analyze the effect of activity on the cooperative motion in our system by computing equal time displacement correlations,
\begin{equation}
C(r) = \frac{\left<u_i u_j\right>-\left<u_i\right> \left<u_j \right>}{\left<u^2\right>},
\end{equation}
where $u$ is the displacement of the particle, i and j denote particle indices, and the angular brackets denote averaging over all particles and several instances. The correlations $C(r)$ are dynamic, they depend on the time scale of observation used for determining the displacements \cite{Dasgupta91,Jack11}. Simulation and experiments have shown that the dynamic correlations are optimal when the time scale of observation equals the relaxation time $\tau_{\alpha}$. So, the relaxation time $\tau_{\alpha}$ obtained from the intermediate scattering function $F_s(k,t)$ (Eq.1) is used to determine the displacements $u_i$. We first discuss the effect of activity on the displacement correlations in the denser system, which is at $\phi\sim0.72$. The Figs.~5(a) and 5(b) are snapshots of the displacement vectors at $T_{eff}=1.0 \textnormal{ and } 4.44$, respectively. The displacements vectors suggest that the motion of particles is correlated over longer distances at $T_{eff}=1$ when compared to the displacements at $T_{eff}=4.4$. To quantify these observations, we compute their correlations $C(r)$ and extract a length scale $\xi$ that characterizes the distance over which the motion is correlated. These results are presented in Fig.~5(c) for $\phi\sim0.72$ and various effective temperatures ranging from $T_{eff}=1 - 4.44$. The correlations $C(r)$ are long-ranged at $T_{eff}=1$ as the correlations reveal a power-law decay extending to $20-30\sigma$. This also confirms that the onset of slow dynamics in the system at higher densities. As the activity increases, the correlations decay faster and they display an exponential decay at $T_{eff}=4.4$. We extract a length scale of the correlated motion by fitting the correlations using a function of the form $C(r) \sim (1/r)~\textnormal{exp}(-r/ \xi) $, where $\xi$ is the correlation length. The best fit method yields length scales $\xi$ that diminishes with increasing activity, which is shown in the inset of Fig.~5c. Clearly, the activity has a strong effect on the length scales of cooperative motion. These observations are similar to the effect of activity on the relaxation time of the system presented in Fig.2a.

\begin{figure}[t!]
\centering
\includegraphics[width=.28\textwidth]{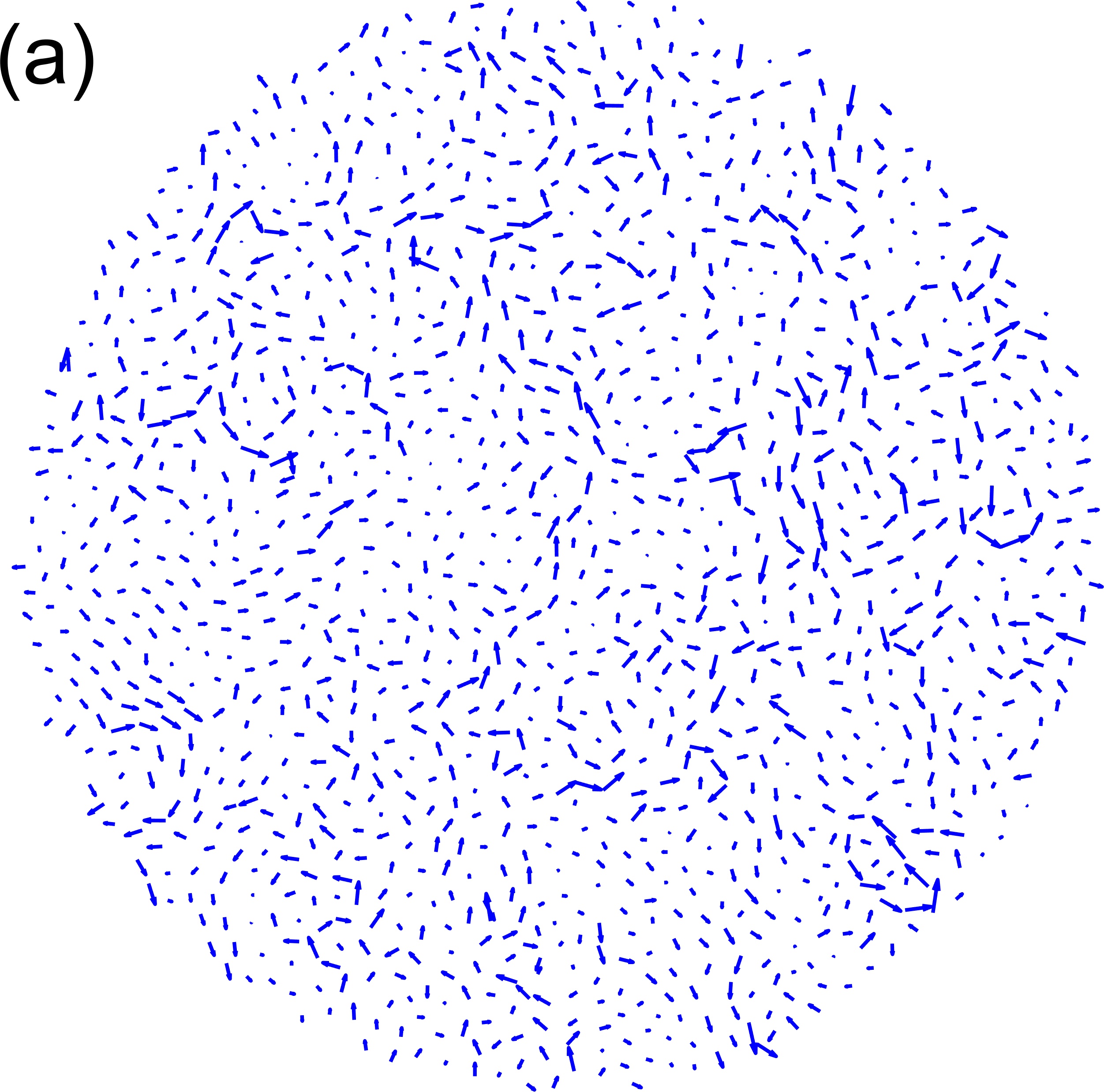}
\includegraphics[width=.28\textwidth]{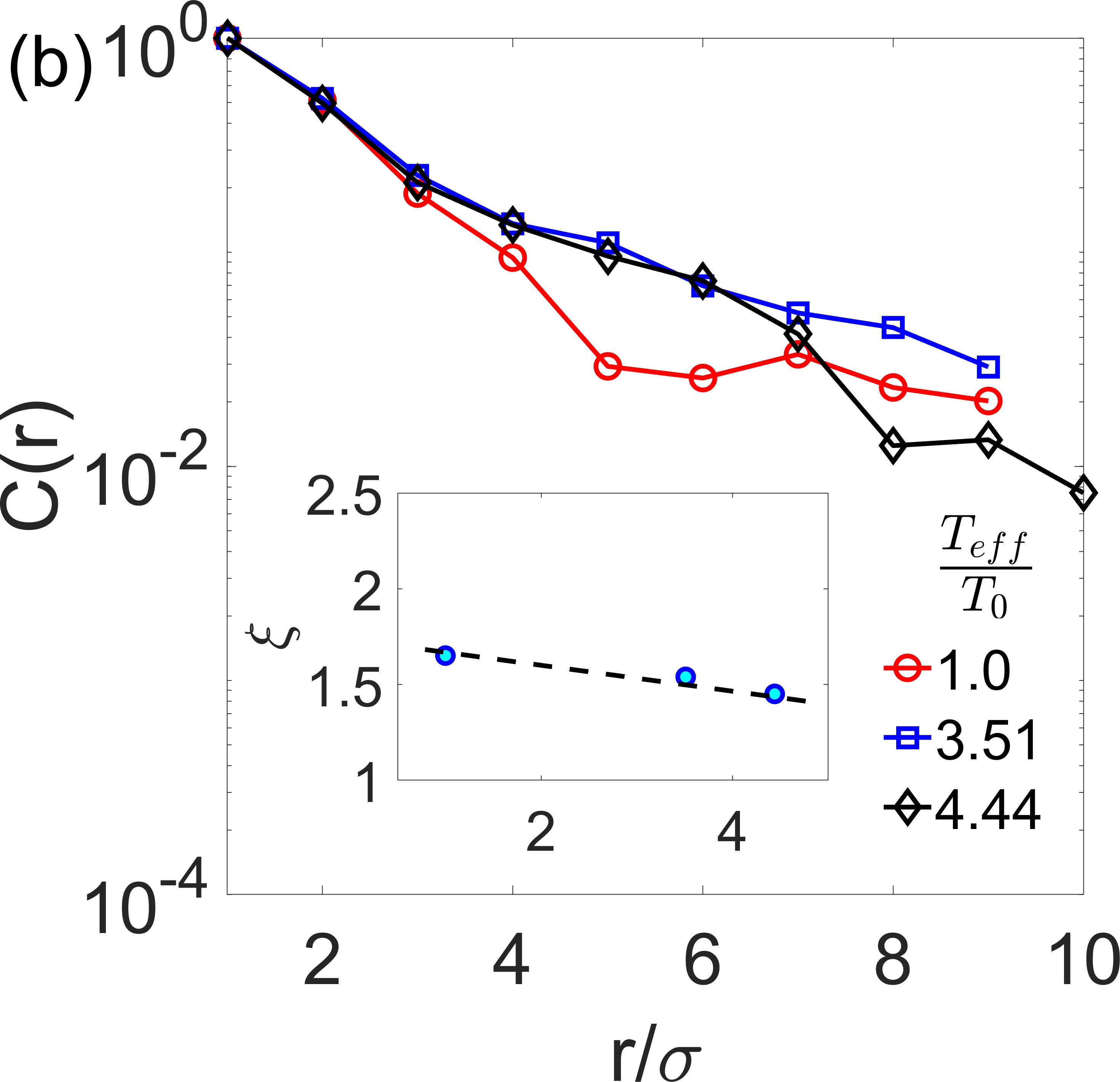}
\caption{Displacement of particles and their spatial correlations at $\phi\sim0.57$. (a) The displacements are computed over a time scale $\tau_{\alpha}$ at $T_{eff}=1$. (c) The displacement correlations C(r) are shown at various effective temperatures $T_{eff}$ in the main panel. The inset shows the correlation lengths $\xi$ extracted from the correlation functions in the main panel.}
\label{Fig6}
\end{figure}

We next focus on the system at a lower area fraction of $\phi\sim0.57$. A snapshot of the displacement vectors of the particles are shown in Fig.6a. Note that the displacements are computed over a time window corresponding to $\tau_{\alpha}$ at $T_{eff}=1$. The motion of the particles appear to be correlated over shorter distances when compared to displacement vectors in Fig.5a. The correlations $C(r)$ in the main panel of Fig.6b indeed display an exponential decay of the form $C(r) \sim \textnormal{exp}(-x/\xi)$. We also observe that increasing the activity does not lead to significant changes in the correlation length. This is consistent with the observation made in Fig.2a, where the activity had a small effect on the relaxation time of the system at $\phi\sim0.57$. The correlation lengths extracted from the best fit method is shown in the inset of Fig.6b. The length scales are considerably small compared to high density systems. 

The analysis presented in this section shows that activity has similar effects on the relaxation timescale and cooperative length scale, both decrease with increase in activity over the range considered in our study.

\section{Cage size and persistence length of active particles}

Recent simulations \cite{Janssen21,Janssen22-2} have pointed out the critical role of cage length and the persistence length of active particles. The authors report an enhancement of diffusivity due to activity when the persistence length, $l_p$, is smaller then the cage size, $l_c$, while it is suppressed when $l_p > l_c$. The central notion is that a particle is efficient in scanning and escaping the cage formed by the neighbors when $l_p < l_c$. In the other limit, ($l_p>>l_c$), it sticks to the edge of the cage, taking longer time to escape. We test these ideas in our experiments by determining $l_p$ and $l_c$. Following the procedure outlined in simulations \cite{Janssen21,Janssen22-2}, the $l_c$ is the average distance of a particle from its nearest neighbors, which is weighted by the pair correlation function $g(r)$. The other lengthscale $l_p$, which is the persistence length of activity, is obtained in the dilute limit. It is obtained from the mean square displacement data, see supplementary information for more details. The Fig.~7 shows the ratio $l_p/l_c$ as the activity of the systems is varied at an area fraction $\phi\sim0.72$. We clearly see that this ratio is of order unity, thus underlining the importance of cage size and persistence length of active particles. 

\begin{figure}[t!]
\centering
\includegraphics[width=.32\textwidth]{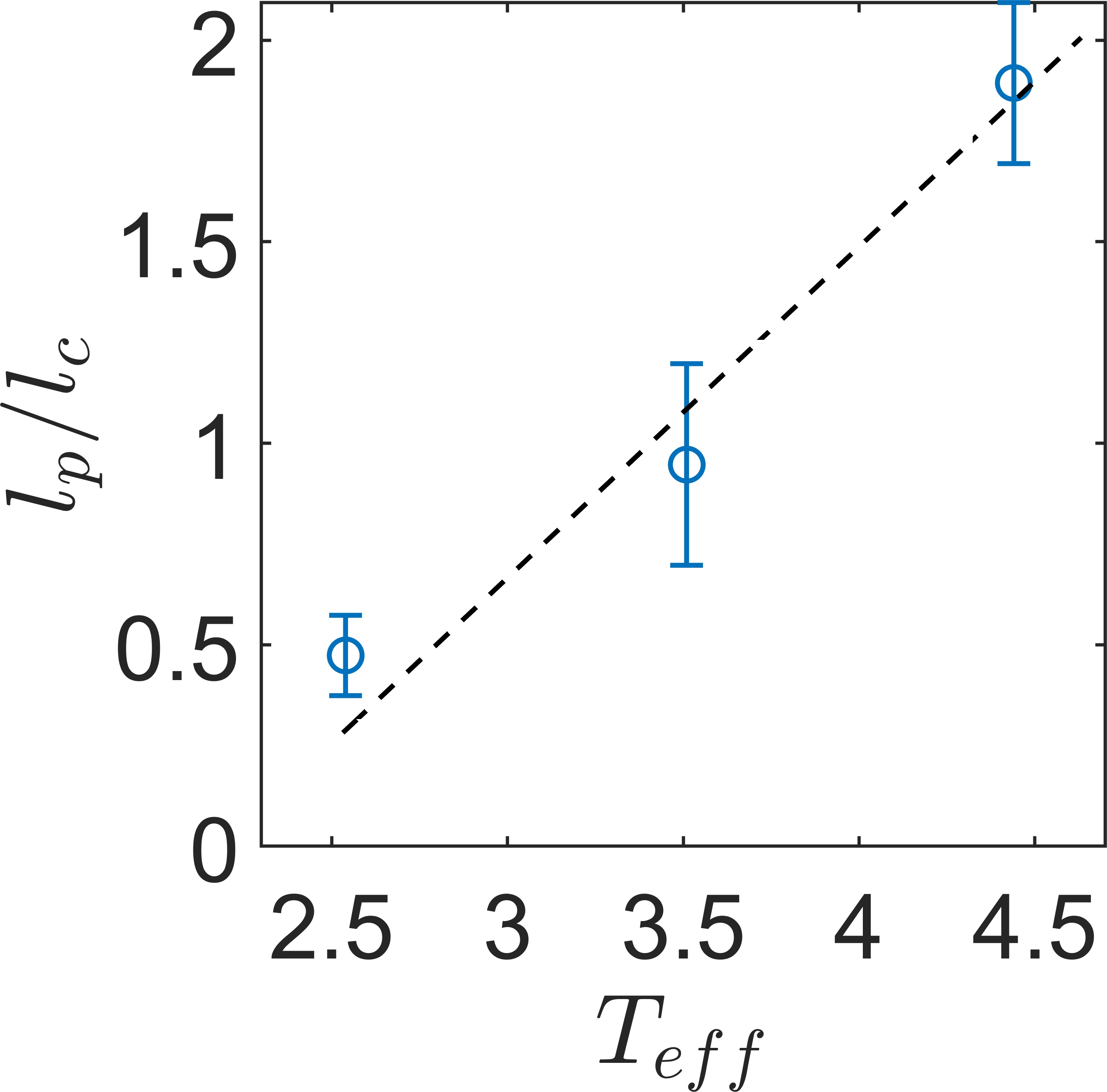}
\caption{The normalised persistence length ($l_p/l_c$) over a range of activities represented by the effective temperature at $\phi\sim0.72$. The dotted line is a guide to the eye.}
\label{Fig7}
\end{figure}

Increasing the effective temperature beyond $T_{eff}=4.44$ leads to destabilisation of the monolayer, the particles do not sediment, especially the small species in our binary mixture. This prevents us from studying interesting features of active solids that manifest at large activities \cite{Karmakar23,Karmakar24}. This can be overcome using larger sized colloidal particles and the measurements are under progress.

\section{Conclusions}
In summary, we present a novel experimental set-up to study the onset of slow dynamics in dense active suspensions containing photoactive Janus colloids. The variation of relaxation timescale with increasing density in Fig.1c and Fig.2b points to a quasi-hard particle nature of active colloids in our system. Besides, these results reveal a monotonic effect of activity on the relaxation of the system across the range of effective temperatures explored in our experiments. Increasing activity leads to faster relaxation at both high and low particle densities. This feature contrasts with the results reported in reference \cite{Leocmach19-1}, where the ergodic and non-ergodic states of the system displayed different relaxation features. The variation of the relaxation timescale ($\tau_{\alpha}$) with the density of our system is well described by the VFT relation, and the fits indicate that the critical density ($\phi_c$) for the glass transition point is pushed to higher densities with increasing activity. These results agree with those found in simulations \cite{Dijkstra13}. The active system exhibits several features of passive thermal systems, such as non-exponential relaxation, cooperative motion, and dynamic arrest.

The effect of activity on the structural ordering of the system is analyzed using two-point density correlations $g(r)$ and local hexagonal ordering $\psi_6$. The peaks of $g(r)$ move to the right with increasing activity, while the local hexagonal ordering $\psi_6$ decreases. Although these changes are small, they point to diminishing structural order due to activity. Furthermore, the displacement correlations reveal a correlation length that decreases with increasing activity. All these observations suggest an enhancement of relaxation due to activity.

Recent reports based on simulations and theory \cite{Karmakar23, Nandi18} suggest a similar effect of activity-driven enhancement of relaxation; however, the dynamic heterogeneity characterized using four-point correlations revealed surprising effects of activity, the relaxation was found to be decoupled from dynamic heterogeneity. Furthermore, simulations \cite{Karmakar24} also reveal non-linear dispersion relation in active systems. Our experimental system motivates further investigations along these directions.




\begin{thebibliography}{48}%
\makeatletter
\providecommand \@ifxundefined [1]{%
 \@ifx{#1\undefined}
}%
\providecommand \@ifnum [1]{%
 \ifnum #1\expandafter \@firstoftwo
 \else \expandafter \@secondoftwo
 \fi
}%
\providecommand \@ifx [1]{%
 \ifx #1\expandafter \@firstoftwo
 \else \expandafter \@secondoftwo
 \fi
}%
\providecommand \natexlab [1]{#1}%
\providecommand \enquote  [1]{``#1''}%
\providecommand \bibnamefont  [1]{#1}%
\providecommand \bibfnamefont [1]{#1}%
\providecommand \citenamefont [1]{#1}%
\providecommand \href@noop [0]{\@secondoftwo}%
\providecommand \href [0]{\begingroup \@sanitize@url \@href}%
\providecommand \@href[1]{\@@startlink{#1}\@@href}%
\providecommand \@@href[1]{\endgroup#1\@@endlink}%
\providecommand \@sanitize@url [0]{\catcode `\\12\catcode `\$12\catcode
  `\&12\catcode `\#12\catcode `\^12\catcode `\_12\catcode `\%12\relax}%
\providecommand \@@startlink[1]{}%
\providecommand \@@endlink[0]{}%
\providecommand \url  [0]{\begingroup\@sanitize@url \@url }%
\providecommand \@url [1]{\endgroup\@href {#1}{\urlprefix }}%
\providecommand \urlprefix  [0]{URL }%
\providecommand \Eprint [0]{\href }%
\providecommand \doibase [0]{https://doi.org/}%
\providecommand \selectlanguage [0]{\@gobble}%
\providecommand \bibinfo  [0]{\@secondoftwo}%
\providecommand \bibfield  [0]{\@secondoftwo}%
\providecommand \translation [1]{[#1]}%
\providecommand \BibitemOpen [0]{}%
\providecommand \bibitemStop [0]{}%
\providecommand \bibitemNoStop [0]{.\EOS\space}%
\providecommand \EOS [0]{\spacefactor3000\relax}%
\providecommand \BibitemShut  [1]{\csname bibitem#1\endcsname}%
\let\auto@bib@innerbib\@empty
\bibitem [{\citenamefont {Debenedetti}\ and\ \citenamefont
  {Stillinger}(2001)}]{Stillinger01}%
  \BibitemOpen
  \bibfield  {author} {\bibinfo {author} {\bibfnamefont {P.~G.}\ \bibnamefont
  {Debenedetti}}\ and\ \bibinfo {author} {\bibfnamefont {F.~H.}\ \bibnamefont
  {Stillinger}},\ }\bibfield  {title} {\bibinfo {title} {Supercooled liquids
  and the glass transition},\ }\href@noop {} {\bibfield  {journal} {\bibinfo
  {journal} {Nature}\ }\textbf {\bibinfo {volume} {410}},\ \bibinfo {pages}
  {259} (\bibinfo {year} {2001})}\BibitemShut {NoStop}%
\bibitem [{\citenamefont {Berthier}\ and\ \citenamefont
  {Biroli}(2011)}]{Biroli11}%
  \BibitemOpen
  \bibfield  {author} {\bibinfo {author} {\bibfnamefont {L.}~\bibnamefont
  {Berthier}}\ and\ \bibinfo {author} {\bibfnamefont {G.}~\bibnamefont
  {Biroli}},\ }\bibfield  {title} {\bibinfo {title} {Theoretical perspective on
  the glass transition and amorphous materials},\ }\href@noop {} {\bibfield
  {journal} {\bibinfo  {journal} {Reviews of modern physics}\ }\textbf
  {\bibinfo {volume} {83}},\ \bibinfo {pages} {587} (\bibinfo {year}
  {2011})}\BibitemShut {NoStop}%
\bibitem [{\citenamefont {Pusey}(1991)}]{Pusey91}%
  \BibitemOpen
  \bibfield  {author} {\bibinfo {author} {\bibfnamefont {P.}~\bibnamefont
  {Pusey}},\ }\bibfield  {title} {\bibinfo {title} {Colloidal suspensions},\
  }in\ \href@noop {} {\emph {\bibinfo {booktitle} {Liquids, Freezing and Glass
  Transition : Les Houches Session LI, 3-28 July, 1989}}},\ \bibinfo {editor}
  {edited by\ \bibinfo {editor} {\bibfnamefont {J.~Z.-J.}\ \bibnamefont
  {J.P.~Hansen}, \bibfnamefont {D.~Levesque}}}\ (\bibinfo  {publisher}
  {North-Holland, Amsterdam},\ \bibinfo {year} {1991})\ pp.\ \bibinfo {pages}
  {199--269}\BibitemShut {NoStop}%
\bibitem [{\citenamefont {Hunter}\ and\ \citenamefont {Weeks}(2012)}]{Weeks12}%
  \BibitemOpen
  \bibfield  {author} {\bibinfo {author} {\bibfnamefont {G.~L.}\ \bibnamefont
  {Hunter}}\ and\ \bibinfo {author} {\bibfnamefont {E.~R.}\ \bibnamefont
  {Weeks}},\ }\bibfield  {title} {\bibinfo {title} {The physics of the
  colloidal glass transition},\ }\href@noop {} {\bibfield  {journal} {\bibinfo
  {journal} {Reports on progress in physics}\ }\textbf {\bibinfo {volume}
  {75}},\ \bibinfo {pages} {066501} (\bibinfo {year} {2012})}\BibitemShut
  {NoStop}%
\bibitem [{\citenamefont {Weeks}\ \emph {et~al.}(2000)\citenamefont {Weeks},
  \citenamefont {Crocker}, \citenamefont {Levitt}, \citenamefont {Schofield},\
  and\ \citenamefont {Weitz}}]{Weitz00}%
  \BibitemOpen
  \bibfield  {author} {\bibinfo {author} {\bibfnamefont {E.~R.}\ \bibnamefont
  {Weeks}}, \bibinfo {author} {\bibfnamefont {J.~C.}\ \bibnamefont {Crocker}},
  \bibinfo {author} {\bibfnamefont {A.~C.}\ \bibnamefont {Levitt}}, \bibinfo
  {author} {\bibfnamefont {A.}~\bibnamefont {Schofield}},\ and\ \bibinfo
  {author} {\bibfnamefont {D.~A.}\ \bibnamefont {Weitz}},\ }\bibfield  {title}
  {\bibinfo {title} {Three-dimensional direct imaging of structural relaxation
  near the colloidal glass transition},\ }\href@noop {} {\bibfield  {journal}
  {\bibinfo  {journal} {Science}\ }\textbf {\bibinfo {volume} {287}},\ \bibinfo
  {pages} {627} (\bibinfo {year} {2000})}\BibitemShut {NoStop}%
\bibitem [{\citenamefont {Pusey}\ and\ \citenamefont
  {Van~Megen}(1986)}]{Megen86}%
  \BibitemOpen
  \bibfield  {author} {\bibinfo {author} {\bibfnamefont {P.~N.}\ \bibnamefont
  {Pusey}}\ and\ \bibinfo {author} {\bibfnamefont {W.}~\bibnamefont
  {Van~Megen}},\ }\bibfield  {title} {\bibinfo {title} {Phase behaviour of
  concentrated suspensions of nearly hard colloidal spheres},\ }\href@noop {}
  {\bibfield  {journal} {\bibinfo  {journal} {Nature}\ }\textbf {\bibinfo
  {volume} {320}},\ \bibinfo {pages} {340} (\bibinfo {year}
  {1986})}\BibitemShut {NoStop}%
\bibitem [{\citenamefont {Marchetti}\ \emph {et~al.}(2013)\citenamefont
  {Marchetti}, \citenamefont {Joanny}, \citenamefont {Ramaswamy}, \citenamefont
  {Liverpool}, \citenamefont {Prost}, \citenamefont {Rao},\ and\ \citenamefont
  {Simha}}]{Simha13}%
  \BibitemOpen
  \bibfield  {author} {\bibinfo {author} {\bibfnamefont {M.~C.}\ \bibnamefont
  {Marchetti}}, \bibinfo {author} {\bibfnamefont {J.-F.}\ \bibnamefont
  {Joanny}}, \bibinfo {author} {\bibfnamefont {S.}~\bibnamefont {Ramaswamy}},
  \bibinfo {author} {\bibfnamefont {T.~B.}\ \bibnamefont {Liverpool}}, \bibinfo
  {author} {\bibfnamefont {J.}~\bibnamefont {Prost}}, \bibinfo {author}
  {\bibfnamefont {M.}~\bibnamefont {Rao}},\ and\ \bibinfo {author}
  {\bibfnamefont {R.~A.}\ \bibnamefont {Simha}},\ }\bibfield  {title} {\bibinfo
  {title} {Hydrodynamics of soft active matter},\ }\href@noop {} {\bibfield
  {journal} {\bibinfo  {journal} {Reviews of modern physics}\ }\textbf
  {\bibinfo {volume} {85}},\ \bibinfo {pages} {1143} (\bibinfo {year}
  {2013})}\BibitemShut {NoStop}%
\bibitem [{\citenamefont {Schoetz}\ \emph {et~al.}(2013)\citenamefont
  {Schoetz}, \citenamefont {Lanio}, \citenamefont {Talbot},\ and\ \citenamefont
  {Manning}}]{Manning13}%
  \BibitemOpen
  \bibfield  {author} {\bibinfo {author} {\bibfnamefont {E.-M.}\ \bibnamefont
  {Schoetz}}, \bibinfo {author} {\bibfnamefont {M.}~\bibnamefont {Lanio}},
  \bibinfo {author} {\bibfnamefont {J.~A.}\ \bibnamefont {Talbot}},\ and\
  \bibinfo {author} {\bibfnamefont {M.~L.}\ \bibnamefont {Manning}},\
  }\bibfield  {title} {\bibinfo {title} {Glassy dynamics in three-dimensional
  embryonic tissues},\ }\href@noop {} {\bibfield  {journal} {\bibinfo
  {journal} {Journal of The Royal Society Interface}\ }\textbf {\bibinfo
  {volume} {10}},\ \bibinfo {pages} {20130726} (\bibinfo {year}
  {2013})}\BibitemShut {NoStop}%
\bibitem [{\citenamefont {Tambe}\ \emph {et~al.}(2011)\citenamefont {Tambe},
  \citenamefont {Corey~Hardin}, \citenamefont {Angelini}, \citenamefont
  {Rajendran}, \citenamefont {Park}, \citenamefont {Serra-Picamal},
  \citenamefont {Zhou}, \citenamefont {Zaman}, \citenamefont {Butler},
  \citenamefont {Weitz} \emph {et~al.}}]{Trepat11}%
  \BibitemOpen
  \bibfield  {author} {\bibinfo {author} {\bibfnamefont {D.~T.}\ \bibnamefont
  {Tambe}}, \bibinfo {author} {\bibfnamefont {C.}~\bibnamefont {Corey~Hardin}},
  \bibinfo {author} {\bibfnamefont {T.~E.}\ \bibnamefont {Angelini}}, \bibinfo
  {author} {\bibfnamefont {K.}~\bibnamefont {Rajendran}}, \bibinfo {author}
  {\bibfnamefont {C.~Y.}\ \bibnamefont {Park}}, \bibinfo {author}
  {\bibfnamefont {X.}~\bibnamefont {Serra-Picamal}}, \bibinfo {author}
  {\bibfnamefont {E.~H.}\ \bibnamefont {Zhou}}, \bibinfo {author}
  {\bibfnamefont {M.~H.}\ \bibnamefont {Zaman}}, \bibinfo {author}
  {\bibfnamefont {J.~P.}\ \bibnamefont {Butler}}, \bibinfo {author}
  {\bibfnamefont {D.~A.}\ \bibnamefont {Weitz}}, \emph {et~al.},\ }\bibfield
  {title} {\bibinfo {title} {Collective cell guidance by cooperative
  intercellular forces},\ }\href@noop {} {\bibfield  {journal} {\bibinfo
  {journal} {Nature materials}\ }\textbf {\bibinfo {volume} {10}},\ \bibinfo
  {pages} {469} (\bibinfo {year} {2011})}\BibitemShut {NoStop}%
\bibitem [{\citenamefont {Oswald}\ \emph {et~al.}(2017)\citenamefont {Oswald},
  \citenamefont {Grosser}, \citenamefont {Smith},\ and\ \citenamefont
  {K{\"a}s}}]{Kas17}%
  \BibitemOpen
  \bibfield  {author} {\bibinfo {author} {\bibfnamefont {L.}~\bibnamefont
  {Oswald}}, \bibinfo {author} {\bibfnamefont {S.}~\bibnamefont {Grosser}},
  \bibinfo {author} {\bibfnamefont {D.~M.}\ \bibnamefont {Smith}},\ and\
  \bibinfo {author} {\bibfnamefont {J.~A.}\ \bibnamefont {K{\"a}s}},\
  }\bibfield  {title} {\bibinfo {title} {Jamming transitions in cancer},\
  }\href@noop {} {\bibfield  {journal} {\bibinfo  {journal} {Journal of physics
  D: Applied physics}\ }\textbf {\bibinfo {volume} {50}},\ \bibinfo {pages}
  {483001} (\bibinfo {year} {2017})}\BibitemShut {NoStop}%
\bibitem [{\citenamefont {Grosser}\ \emph {et~al.}(2021)\citenamefont
  {Grosser}, \citenamefont {Lippoldt}, \citenamefont {Oswald}, \citenamefont
  {Merkel}, \citenamefont {Sussman}, \citenamefont {Renner}, \citenamefont
  {Gottheil}, \citenamefont {Morawetz}, \citenamefont {Fuhs}, \citenamefont
  {Xie} \emph {et~al.}}]{Kas21}%
  \BibitemOpen
  \bibfield  {author} {\bibinfo {author} {\bibfnamefont {S.}~\bibnamefont
  {Grosser}}, \bibinfo {author} {\bibfnamefont {J.}~\bibnamefont {Lippoldt}},
  \bibinfo {author} {\bibfnamefont {L.}~\bibnamefont {Oswald}}, \bibinfo
  {author} {\bibfnamefont {M.}~\bibnamefont {Merkel}}, \bibinfo {author}
  {\bibfnamefont {D.~M.}\ \bibnamefont {Sussman}}, \bibinfo {author}
  {\bibfnamefont {F.}~\bibnamefont {Renner}}, \bibinfo {author} {\bibfnamefont
  {P.}~\bibnamefont {Gottheil}}, \bibinfo {author} {\bibfnamefont {E.~W.}\
  \bibnamefont {Morawetz}}, \bibinfo {author} {\bibfnamefont {T.}~\bibnamefont
  {Fuhs}}, \bibinfo {author} {\bibfnamefont {X.}~\bibnamefont {Xie}}, \emph
  {et~al.},\ }\bibfield  {title} {\bibinfo {title} {Cell and nucleus shape as
  an indicator of tissue fluidity in carcinoma},\ }\href@noop {} {\bibfield
  {journal} {\bibinfo  {journal} {Physical Review X}\ }\textbf {\bibinfo
  {volume} {11}},\ \bibinfo {pages} {011033} (\bibinfo {year}
  {2021})}\BibitemShut {NoStop}%
\bibitem [{\citenamefont {Parry}\ \emph {et~al.}(2014)\citenamefont {Parry},
  \citenamefont {Surovtsev}, \citenamefont {Cabeen}, \citenamefont {O’Hern},
  \citenamefont {Dufresne},\ and\ \citenamefont
  {Jacobs-Wagner}}]{Jacobs-Wagner14}%
  \BibitemOpen
  \bibfield  {author} {\bibinfo {author} {\bibfnamefont {B.~R.}\ \bibnamefont
  {Parry}}, \bibinfo {author} {\bibfnamefont {I.~V.}\ \bibnamefont
  {Surovtsev}}, \bibinfo {author} {\bibfnamefont {M.~T.}\ \bibnamefont
  {Cabeen}}, \bibinfo {author} {\bibfnamefont {C.~S.}\ \bibnamefont
  {O’Hern}}, \bibinfo {author} {\bibfnamefont {E.~R.}\ \bibnamefont
  {Dufresne}},\ and\ \bibinfo {author} {\bibfnamefont {C.}~\bibnamefont
  {Jacobs-Wagner}},\ }\bibfield  {title} {\bibinfo {title} {The bacterial
  cytoplasm has glass-like properties and is fluidized by metabolic activity},\
  }\href@noop {} {\bibfield  {journal} {\bibinfo  {journal} {Cell}\ }\textbf
  {\bibinfo {volume} {156}},\ \bibinfo {pages} {183} (\bibinfo {year}
  {2014})}\BibitemShut {NoStop}%
\bibitem [{\citenamefont {Berthier}\ and\ \citenamefont
  {Kurchan}(2013)}]{Kurchan13}%
  \BibitemOpen
  \bibfield  {author} {\bibinfo {author} {\bibfnamefont {L.}~\bibnamefont
  {Berthier}}\ and\ \bibinfo {author} {\bibfnamefont {J.}~\bibnamefont
  {Kurchan}},\ }\bibfield  {title} {\bibinfo {title} {Non-equilibrium glass
  transitions in driven and active matter},\ }\href@noop {} {\bibfield
  {journal} {\bibinfo  {journal} {Nature Physics}\ }\textbf {\bibinfo {volume}
  {9}},\ \bibinfo {pages} {310} (\bibinfo {year} {2013})}\BibitemShut {NoStop}%
\bibitem [{\citenamefont {Buttinoni}\ \emph {et~al.}(2012)\citenamefont
  {Buttinoni}, \citenamefont {Volpe}, \citenamefont {K{\"u}mmel}, \citenamefont
  {Volpe},\ and\ \citenamefont {Bechinger}}]{Bechinger12}%
  \BibitemOpen
  \bibfield  {author} {\bibinfo {author} {\bibfnamefont {I.}~\bibnamefont
  {Buttinoni}}, \bibinfo {author} {\bibfnamefont {G.}~\bibnamefont {Volpe}},
  \bibinfo {author} {\bibfnamefont {F.}~\bibnamefont {K{\"u}mmel}}, \bibinfo
  {author} {\bibfnamefont {G.}~\bibnamefont {Volpe}},\ and\ \bibinfo {author}
  {\bibfnamefont {C.}~\bibnamefont {Bechinger}},\ }\bibfield  {title} {\bibinfo
  {title} {Active brownian motion tunable by light},\ }\href@noop {} {\bibfield
   {journal} {\bibinfo  {journal} {Journal of Physics: Condensed Matter}\
  }\textbf {\bibinfo {volume} {24}},\ \bibinfo {pages} {284129} (\bibinfo
  {year} {2012})}\BibitemShut {NoStop}%
\bibitem [{\citenamefont {Janssen}(2019)}]{Janssen19}%
  \BibitemOpen
  \bibfield  {author} {\bibinfo {author} {\bibfnamefont {L.~M.}\ \bibnamefont
  {Janssen}},\ }\bibfield  {title} {\bibinfo {title} {Active glasses},\
  }\href@noop {} {\bibfield  {journal} {\bibinfo  {journal} {Journal of
  Physics: Condensed Matter}\ }\textbf {\bibinfo {volume} {31}},\ \bibinfo
  {pages} {503002} (\bibinfo {year} {2019})}\BibitemShut {NoStop}%
\bibitem [{\citenamefont {Fily}\ and\ \citenamefont
  {Marchetti}(2012)}]{Marchetti12}%
  \BibitemOpen
  \bibfield  {author} {\bibinfo {author} {\bibfnamefont {Y.}~\bibnamefont
  {Fily}}\ and\ \bibinfo {author} {\bibfnamefont {M.~C.}\ \bibnamefont
  {Marchetti}},\ }\bibfield  {title} {\bibinfo {title} {Athermal phase
  separation of self-propelled particles with no alignment},\ }\href@noop {}
  {\bibfield  {journal} {\bibinfo  {journal} {Physical review letters}\
  }\textbf {\bibinfo {volume} {108}},\ \bibinfo {pages} {235702} (\bibinfo
  {year} {2012})}\BibitemShut {NoStop}%
\bibitem [{\citenamefont {Romanczuk}\ \emph {et~al.}(2012)\citenamefont
  {Romanczuk}, \citenamefont {B{\"a}r}, \citenamefont {Ebeling}, \citenamefont
  {Lindner},\ and\ \citenamefont {Schimansky-Geier}}]{Schimansky-Geier12}%
  \BibitemOpen
  \bibfield  {author} {\bibinfo {author} {\bibfnamefont {P.}~\bibnamefont
  {Romanczuk}}, \bibinfo {author} {\bibfnamefont {M.}~\bibnamefont {B{\"a}r}},
  \bibinfo {author} {\bibfnamefont {W.}~\bibnamefont {Ebeling}}, \bibinfo
  {author} {\bibfnamefont {B.}~\bibnamefont {Lindner}},\ and\ \bibinfo {author}
  {\bibfnamefont {L.}~\bibnamefont {Schimansky-Geier}},\ }\bibfield  {title}
  {\bibinfo {title} {Active brownian particles: From individual to collective
  stochastic dynamics},\ }\href@noop {} {\bibfield  {journal} {\bibinfo
  {journal} {The European Physical Journal Special Topics}\ }\textbf {\bibinfo
  {volume} {202}},\ \bibinfo {pages} {1} (\bibinfo {year} {2012})}\BibitemShut
  {NoStop}%
\bibitem [{\citenamefont {Szamel}(2014)}]{Szamel14}%
  \BibitemOpen
  \bibfield  {author} {\bibinfo {author} {\bibfnamefont {G.}~\bibnamefont
  {Szamel}},\ }\bibfield  {title} {\bibinfo {title} {Self-propelled particle in
  an external potential: Existence of an effective temperature},\ }\href@noop
  {} {\bibfield  {journal} {\bibinfo  {journal} {Physical Review E}\ }\textbf
  {\bibinfo {volume} {90}},\ \bibinfo {pages} {012111} (\bibinfo {year}
  {2014})}\BibitemShut {NoStop}%
\bibitem [{\citenamefont {Maggi}\ \emph {et~al.}(2015)\citenamefont {Maggi},
  \citenamefont {Marconi}, \citenamefont {Gnan},\ and\ \citenamefont
  {Di~Leonardo}}]{Leonardo15}%
  \BibitemOpen
  \bibfield  {author} {\bibinfo {author} {\bibfnamefont {C.}~\bibnamefont
  {Maggi}}, \bibinfo {author} {\bibfnamefont {U.~M.~B.}\ \bibnamefont
  {Marconi}}, \bibinfo {author} {\bibfnamefont {N.}~\bibnamefont {Gnan}},\ and\
  \bibinfo {author} {\bibfnamefont {R.}~\bibnamefont {Di~Leonardo}},\
  }\bibfield  {title} {\bibinfo {title} {Multidimensional stationary
  probability distribution for interacting active particles},\ }\href@noop {}
  {\bibfield  {journal} {\bibinfo  {journal} {Scientific reports}\ }\textbf
  {\bibinfo {volume} {5}},\ \bibinfo {pages} {10742} (\bibinfo {year}
  {2015})}\BibitemShut {NoStop}%
\bibitem [{\citenamefont {Howse}\ \emph {et~al.}(2007)\citenamefont {Howse},
  \citenamefont {Jones}, \citenamefont {Ryan}, \citenamefont {Gough},
  \citenamefont {Vafabakhsh},\ and\ \citenamefont
  {Golestanian}}]{Golestanian07}%
  \BibitemOpen
  \bibfield  {author} {\bibinfo {author} {\bibfnamefont {J.~R.}\ \bibnamefont
  {Howse}}, \bibinfo {author} {\bibfnamefont {R.~A.}\ \bibnamefont {Jones}},
  \bibinfo {author} {\bibfnamefont {A.~J.}\ \bibnamefont {Ryan}}, \bibinfo
  {author} {\bibfnamefont {T.}~\bibnamefont {Gough}}, \bibinfo {author}
  {\bibfnamefont {R.}~\bibnamefont {Vafabakhsh}},\ and\ \bibinfo {author}
  {\bibfnamefont {R.}~\bibnamefont {Golestanian}},\ }\bibfield  {title}
  {\bibinfo {title} {Self-motile colloidal particles: from directed propulsion
  to random walk},\ }\href@noop {} {\bibfield  {journal} {\bibinfo  {journal}
  {Physical review letters}\ }\textbf {\bibinfo {volume} {99}},\ \bibinfo
  {pages} {048102} (\bibinfo {year} {2007})}\BibitemShut {NoStop}%
\bibitem [{\citenamefont {Theurkauff}\ \emph {et~al.}(2012)\citenamefont
  {Theurkauff}, \citenamefont {Cottin-Bizonne}, \citenamefont {Palacci},
  \citenamefont {Ybert},\ and\ \citenamefont {Bocquet}}]{Bocquet12}%
  \BibitemOpen
  \bibfield  {author} {\bibinfo {author} {\bibfnamefont {I.}~\bibnamefont
  {Theurkauff}}, \bibinfo {author} {\bibfnamefont {C.}~\bibnamefont
  {Cottin-Bizonne}}, \bibinfo {author} {\bibfnamefont {J.}~\bibnamefont
  {Palacci}}, \bibinfo {author} {\bibfnamefont {C.}~\bibnamefont {Ybert}},\
  and\ \bibinfo {author} {\bibfnamefont {L.}~\bibnamefont {Bocquet}},\
  }\bibfield  {title} {\bibinfo {title} {Dynamic clustering in active colloidal
  suspensions with chemical signaling},\ }\href
  {https://doi.org/10.1103/PhysRevLett.108.268303} {\bibfield  {journal}
  {\bibinfo  {journal} {Phys. Rev. Lett.}\ }\textbf {\bibinfo {volume} {108}},\
  \bibinfo {pages} {268303} (\bibinfo {year} {2012})}\BibitemShut {NoStop}%
\bibitem [{\citenamefont {Bechinger}\ \emph {et~al.}(2016)\citenamefont
  {Bechinger}, \citenamefont {Di~Leonardo}, \citenamefont {L{\"o}wen},
  \citenamefont {Reichhardt}, \citenamefont {Volpe},\ and\ \citenamefont
  {Volpe}}]{Volpe16}%
  \BibitemOpen
  \bibfield  {author} {\bibinfo {author} {\bibfnamefont {C.}~\bibnamefont
  {Bechinger}}, \bibinfo {author} {\bibfnamefont {R.}~\bibnamefont
  {Di~Leonardo}}, \bibinfo {author} {\bibfnamefont {H.}~\bibnamefont
  {L{\"o}wen}}, \bibinfo {author} {\bibfnamefont {C.}~\bibnamefont
  {Reichhardt}}, \bibinfo {author} {\bibfnamefont {G.}~\bibnamefont {Volpe}},\
  and\ \bibinfo {author} {\bibfnamefont {G.}~\bibnamefont {Volpe}},\ }\bibfield
   {title} {\bibinfo {title} {Active particles in complex and crowded
  environments},\ }\href@noop {} {\bibfield  {journal} {\bibinfo  {journal}
  {Reviews of Modern Physics}\ }\textbf {\bibinfo {volume} {88}},\ \bibinfo
  {pages} {045006} (\bibinfo {year} {2016})}\BibitemShut {NoStop}%
\bibitem [{\citenamefont {Ni}\ \emph {et~al.}(2013)\citenamefont {Ni},
  \citenamefont {Stuart},\ and\ \citenamefont {Dijkstra}}]{Dijkstra13}%
  \BibitemOpen
  \bibfield  {author} {\bibinfo {author} {\bibfnamefont {R.}~\bibnamefont
  {Ni}}, \bibinfo {author} {\bibfnamefont {M.~A.~C.}\ \bibnamefont {Stuart}},\
  and\ \bibinfo {author} {\bibfnamefont {M.}~\bibnamefont {Dijkstra}},\
  }\bibfield  {title} {\bibinfo {title} {Pushing the glass transition towards
  random close packing using self-propelled hard spheres},\ }\href@noop {}
  {\bibfield  {journal} {\bibinfo  {journal} {Nature communications}\ }\textbf
  {\bibinfo {volume} {4}},\ \bibinfo {pages} {2704} (\bibinfo {year}
  {2013})}\BibitemShut {NoStop}%
\bibitem [{\citenamefont {Berthier}(2014)}]{Berthier14}%
  \BibitemOpen
  \bibfield  {author} {\bibinfo {author} {\bibfnamefont {L.}~\bibnamefont
  {Berthier}},\ }\bibfield  {title} {\bibinfo {title} {Nonequilibrium glassy
  dynamics of self-propelled hard disks},\ }\href@noop {} {\bibfield  {journal}
  {\bibinfo  {journal} {Physical review letters}\ }\textbf {\bibinfo {volume}
  {112}},\ \bibinfo {pages} {220602} (\bibinfo {year} {2014})}\BibitemShut
  {NoStop}%
\bibitem [{\citenamefont {Szamel}\ \emph {et~al.}(2015)\citenamefont {Szamel},
  \citenamefont {Flenner},\ and\ \citenamefont {Berthier}}]{Berthier15}%
  \BibitemOpen
  \bibfield  {author} {\bibinfo {author} {\bibfnamefont {G.}~\bibnamefont
  {Szamel}}, \bibinfo {author} {\bibfnamefont {E.}~\bibnamefont {Flenner}},\
  and\ \bibinfo {author} {\bibfnamefont {L.}~\bibnamefont {Berthier}},\
  }\bibfield  {title} {\bibinfo {title} {Glassy dynamics of athermal
  self-propelled particles: Computer simulations and a nonequilibrium
  microscopic theory},\ }\href@noop {} {\bibfield  {journal} {\bibinfo
  {journal} {Physical Review E}\ }\textbf {\bibinfo {volume} {91}},\ \bibinfo
  {pages} {062304} (\bibinfo {year} {2015})}\BibitemShut {NoStop}%
\bibitem [{\citenamefont {Mandal}\ \emph {et~al.}(2016)\citenamefont {Mandal},
  \citenamefont {Bhuyan}, \citenamefont {Rao},\ and\ \citenamefont
  {Dasgupta}}]{Dasgupta16}%
  \BibitemOpen
  \bibfield  {author} {\bibinfo {author} {\bibfnamefont {R.}~\bibnamefont
  {Mandal}}, \bibinfo {author} {\bibfnamefont {P.~J.}\ \bibnamefont {Bhuyan}},
  \bibinfo {author} {\bibfnamefont {M.}~\bibnamefont {Rao}},\ and\ \bibinfo
  {author} {\bibfnamefont {C.}~\bibnamefont {Dasgupta}},\ }\bibfield  {title}
  {\bibinfo {title} {Active fluidization in dense glassy systems},\ }\href@noop
  {} {\bibfield  {journal} {\bibinfo  {journal} {Soft Matter}\ }\textbf
  {\bibinfo {volume} {12}},\ \bibinfo {pages} {6268} (\bibinfo {year}
  {2016})}\BibitemShut {NoStop}%
\bibitem [{\citenamefont {Flenner}\ \emph {et~al.}(2016)\citenamefont
  {Flenner}, \citenamefont {Szamel},\ and\ \citenamefont
  {Berthier}}]{Berthier16}%
  \BibitemOpen
  \bibfield  {author} {\bibinfo {author} {\bibfnamefont {E.}~\bibnamefont
  {Flenner}}, \bibinfo {author} {\bibfnamefont {G.}~\bibnamefont {Szamel}},\
  and\ \bibinfo {author} {\bibfnamefont {L.}~\bibnamefont {Berthier}},\
  }\bibfield  {title} {\bibinfo {title} {The nonequilibrium glassy dynamics of
  self-propelled particles},\ }\href@noop {} {\bibfield  {journal} {\bibinfo
  {journal} {Soft matter}\ }\textbf {\bibinfo {volume} {12}},\ \bibinfo {pages}
  {7136} (\bibinfo {year} {2016})}\BibitemShut {NoStop}%
\bibitem [{\citenamefont {Berthier}\ \emph {et~al.}(2017)\citenamefont
  {Berthier}, \citenamefont {Flenner},\ and\ \citenamefont
  {Szamel}}]{Szamel17}%
  \BibitemOpen
  \bibfield  {author} {\bibinfo {author} {\bibfnamefont {L.}~\bibnamefont
  {Berthier}}, \bibinfo {author} {\bibfnamefont {E.}~\bibnamefont {Flenner}},\
  and\ \bibinfo {author} {\bibfnamefont {G.}~\bibnamefont {Szamel}},\
  }\bibfield  {title} {\bibinfo {title} {How active forces influence
  nonequilibrium glass transitions},\ }\href@noop {} {\bibfield  {journal}
  {\bibinfo  {journal} {New Journal of Physics}\ }\textbf {\bibinfo {volume}
  {19}},\ \bibinfo {pages} {125006} (\bibinfo {year} {2017})}\BibitemShut
  {NoStop}%
\bibitem [{\citenamefont {Debets}\ \emph {et~al.}(2021)\citenamefont {Debets},
  \citenamefont {De~Wit},\ and\ \citenamefont {Janssen}}]{Janssen21}%
  \BibitemOpen
  \bibfield  {author} {\bibinfo {author} {\bibfnamefont {V.~E.}\ \bibnamefont
  {Debets}}, \bibinfo {author} {\bibfnamefont {X.~M.}\ \bibnamefont {De~Wit}},\
  and\ \bibinfo {author} {\bibfnamefont {L.~M.}\ \bibnamefont {Janssen}},\
  }\bibfield  {title} {\bibinfo {title} {Cage length controls the nonmonotonic
  dynamics of active glassy matter},\ }\href@noop {} {\bibfield  {journal}
  {\bibinfo  {journal} {Physical Review Letters}\ }\textbf {\bibinfo {volume}
  {127}},\ \bibinfo {pages} {278002} (\bibinfo {year} {2021})}\BibitemShut
  {NoStop}%
\bibitem [{\citenamefont {Mandal}\ and\ \citenamefont
  {Sollich}(2020)}]{Sollich20}%
  \BibitemOpen
  \bibfield  {author} {\bibinfo {author} {\bibfnamefont {R.}~\bibnamefont
  {Mandal}}\ and\ \bibinfo {author} {\bibfnamefont {P.}~\bibnamefont
  {Sollich}},\ }\bibfield  {title} {\bibinfo {title} {Multiple types of aging
  in active glasses},\ }\href@noop {} {\bibfield  {journal} {\bibinfo
  {journal} {Physical Review Letters}\ }\textbf {\bibinfo {volume} {125}},\
  \bibinfo {pages} {218001} (\bibinfo {year} {2020})}\BibitemShut {NoStop}%
\bibitem [{\citenamefont {Janzen}\ and\ \citenamefont
  {Janssen}(2022)}]{Janssen22}%
  \BibitemOpen
  \bibfield  {author} {\bibinfo {author} {\bibfnamefont {G.}~\bibnamefont
  {Janzen}}\ and\ \bibinfo {author} {\bibfnamefont {L.~M.}\ \bibnamefont
  {Janssen}},\ }\bibfield  {title} {\bibinfo {title} {Aging in thermal active
  glasses},\ }\href@noop {} {\bibfield  {journal} {\bibinfo  {journal}
  {Physical Review Research}\ }\textbf {\bibinfo {volume} {4}},\ \bibinfo
  {pages} {L012038} (\bibinfo {year} {2022})}\BibitemShut {NoStop}%
\bibitem [{\citenamefont {Arora}\ \emph {et~al.}(2022)\citenamefont {Arora},
  \citenamefont {Sood},\ and\ \citenamefont {Ganapathy}}]{Ganapathy22}%
  \BibitemOpen
  \bibfield  {author} {\bibinfo {author} {\bibfnamefont {P.}~\bibnamefont
  {Arora}}, \bibinfo {author} {\bibfnamefont {A.}~\bibnamefont {Sood}},\ and\
  \bibinfo {author} {\bibfnamefont {R.}~\bibnamefont {Ganapathy}},\ }\bibfield
  {title} {\bibinfo {title} {Motile topological defects hinder dynamical arrest
  in dense liquids of active ellipsoids},\ }\href@noop {} {\bibfield  {journal}
  {\bibinfo  {journal} {Physical Review Letters}\ }\textbf {\bibinfo {volume}
  {128}},\ \bibinfo {pages} {178002} (\bibinfo {year} {2022})}\BibitemShut
  {NoStop}%
\bibitem [{\citenamefont {Klongvessa}\ \emph
  {et~al.}(2019{\natexlab{a}})\citenamefont {Klongvessa}, \citenamefont
  {Ginot}, \citenamefont {Ybert}, \citenamefont {Cottin-Bizonne},\ and\
  \citenamefont {Leocmach}}]{Leocmach19-1}%
  \BibitemOpen
  \bibfield  {author} {\bibinfo {author} {\bibfnamefont {N.}~\bibnamefont
  {Klongvessa}}, \bibinfo {author} {\bibfnamefont {F.}~\bibnamefont {Ginot}},
  \bibinfo {author} {\bibfnamefont {C.}~\bibnamefont {Ybert}}, \bibinfo
  {author} {\bibfnamefont {C.}~\bibnamefont {Cottin-Bizonne}},\ and\ \bibinfo
  {author} {\bibfnamefont {M.}~\bibnamefont {Leocmach}},\ }\bibfield  {title}
  {\bibinfo {title} {Active glass: Ergodicity breaking dramatically affects
  response to self-propulsion},\ }\href@noop {} {\bibfield  {journal} {\bibinfo
   {journal} {Physical review letters}\ }\textbf {\bibinfo {volume} {123}},\
  \bibinfo {pages} {248004} (\bibinfo {year} {2019}{\natexlab{a}})}\BibitemShut
  {NoStop}%
\bibitem [{\citenamefont {Klongvessa}\ \emph
  {et~al.}(2019{\natexlab{b}})\citenamefont {Klongvessa}, \citenamefont
  {Ginot}, \citenamefont {Ybert}, \citenamefont {Cottin-Bizonne},\ and\
  \citenamefont {Leocmach}}]{Leocmach19-2}%
  \BibitemOpen
  \bibfield  {author} {\bibinfo {author} {\bibfnamefont {N.}~\bibnamefont
  {Klongvessa}}, \bibinfo {author} {\bibfnamefont {F.}~\bibnamefont {Ginot}},
  \bibinfo {author} {\bibfnamefont {C.}~\bibnamefont {Ybert}}, \bibinfo
  {author} {\bibfnamefont {C.}~\bibnamefont {Cottin-Bizonne}},\ and\ \bibinfo
  {author} {\bibfnamefont {M.}~\bibnamefont {Leocmach}},\ }\bibfield  {title}
  {\bibinfo {title} {Nonmonotonic behavior in dense assemblies of active
  colloids},\ }\href@noop {} {\bibfield  {journal} {\bibinfo  {journal}
  {Physical Review E}\ }\textbf {\bibinfo {volume} {100}},\ \bibinfo {pages}
  {062603} (\bibinfo {year} {2019}{\natexlab{b}})}\BibitemShut {NoStop}%
\bibitem [{\citenamefont {Philippe}\ \emph {et~al.}(2018)\citenamefont
  {Philippe}, \citenamefont {Truzzolillo}, \citenamefont {Galvan-Myoshi},
  \citenamefont {Dieudonn{\'e}-George}, \citenamefont {Trappe}, \citenamefont
  {Berthier},\ and\ \citenamefont {Cipelletti}}]{Cipelletti18}%
  \BibitemOpen
  \bibfield  {author} {\bibinfo {author} {\bibfnamefont {A.-M.}\ \bibnamefont
  {Philippe}}, \bibinfo {author} {\bibfnamefont {D.}~\bibnamefont
  {Truzzolillo}}, \bibinfo {author} {\bibfnamefont {J.}~\bibnamefont
  {Galvan-Myoshi}}, \bibinfo {author} {\bibfnamefont {P.}~\bibnamefont
  {Dieudonn{\'e}-George}}, \bibinfo {author} {\bibfnamefont {V.}~\bibnamefont
  {Trappe}}, \bibinfo {author} {\bibfnamefont {L.}~\bibnamefont {Berthier}},\
  and\ \bibinfo {author} {\bibfnamefont {L.}~\bibnamefont {Cipelletti}},\
  }\bibfield  {title} {\bibinfo {title} {Glass transition of soft colloids},\
  }\href@noop {} {\bibfield  {journal} {\bibinfo  {journal} {Physical Review
  E}\ }\textbf {\bibinfo {volume} {97}},\ \bibinfo {pages} {040601} (\bibinfo
  {year} {2018})}\BibitemShut {NoStop}%
\bibitem [{\citenamefont {Debets}\ and\ \citenamefont
  {Janssen}(2022)}]{Janssen22-2}%
  \BibitemOpen
  \bibfield  {author} {\bibinfo {author} {\bibfnamefont {V.~E.}\ \bibnamefont
  {Debets}}\ and\ \bibinfo {author} {\bibfnamefont {L.~M.}\ \bibnamefont
  {Janssen}},\ }\bibfield  {title} {\bibinfo {title} {Influence of particle
  softness on active glassy dynamics},\ }\href@noop {} {\bibfield  {journal}
  {\bibinfo  {journal} {Physical Review Research}\ }\textbf {\bibinfo {volume}
  {4}},\ \bibinfo {pages} {L042033} (\bibinfo {year} {2022})}\BibitemShut
  {NoStop}%
\bibitem [{\citenamefont {Singh}\ \emph {et~al.}(2017)\citenamefont {Singh},
  \citenamefont {Choudhury}, \citenamefont {Fischer},\ and\ \citenamefont
  {Mark}}]{Singh17}%
  \BibitemOpen
  \bibfield  {author} {\bibinfo {author} {\bibfnamefont {D.~P.}\ \bibnamefont
  {Singh}}, \bibinfo {author} {\bibfnamefont {U.}~\bibnamefont {Choudhury}},
  \bibinfo {author} {\bibfnamefont {P.}~\bibnamefont {Fischer}},\ and\ \bibinfo
  {author} {\bibfnamefont {A.~G.}\ \bibnamefont {Mark}},\ }\bibfield  {title}
  {\bibinfo {title} {Non-equilibrium assembly of light-activated colloidal
  mixtures},\ }\href@noop {} {\bibfield  {journal} {\bibinfo  {journal}
  {Advanced Materials}\ }\textbf {\bibinfo {volume} {29}},\ \bibinfo {pages}
  {1701328} (\bibinfo {year} {2017})}\BibitemShut {NoStop}%
\bibitem [{\citenamefont {Hong}\ \emph {et~al.}(2010)\citenamefont {Hong},
  \citenamefont {Diaz}, \citenamefont {C{\'o}rdova-Figueroa},\ and\
  \citenamefont {Sen}}]{Sen10}%
  \BibitemOpen
  \bibfield  {author} {\bibinfo {author} {\bibfnamefont {Y.}~\bibnamefont
  {Hong}}, \bibinfo {author} {\bibfnamefont {M.}~\bibnamefont {Diaz}}, \bibinfo
  {author} {\bibfnamefont {U.~M.}\ \bibnamefont {C{\'o}rdova-Figueroa}},\ and\
  \bibinfo {author} {\bibfnamefont {A.}~\bibnamefont {Sen}},\ }\bibfield
  {title} {\bibinfo {title} {Light-driven titanium-dioxide-based reversible
  microfireworks and micromotor/micropump systems},\ }\href@noop {} {\bibfield
  {journal} {\bibinfo  {journal} {Advanced Functional Materials}\ }\textbf
  {\bibinfo {volume} {20}},\ \bibinfo {pages} {1568} (\bibinfo {year}
  {2010})}\BibitemShut {NoStop}%
\bibitem [{\citenamefont {Vivek}\ \emph {et~al.}(2017)\citenamefont {Vivek},
  \citenamefont {Kelleher}, \citenamefont {Chaikin},\ and\ \citenamefont
  {Weeks}}]{Weeks17}%
  \BibitemOpen
  \bibfield  {author} {\bibinfo {author} {\bibfnamefont {S.}~\bibnamefont
  {Vivek}}, \bibinfo {author} {\bibfnamefont {C.~P.}\ \bibnamefont {Kelleher}},
  \bibinfo {author} {\bibfnamefont {P.~M.}\ \bibnamefont {Chaikin}},\ and\
  \bibinfo {author} {\bibfnamefont {E.~R.}\ \bibnamefont {Weeks}},\ }\bibfield
  {title} {\bibinfo {title} {Long-wavelength fluctuations and the glass
  transition in two dimensions and three dimensions},\ }\href@noop {}
  {\bibfield  {journal} {\bibinfo  {journal} {Proceedings of the National
  Academy of Sciences}\ }\textbf {\bibinfo {volume} {114}},\ \bibinfo {pages}
  {1850} (\bibinfo {year} {2017})}\BibitemShut {NoStop}%
\bibitem [{\citenamefont {Illing}\ \emph {et~al.}(2017)\citenamefont {Illing},
  \citenamefont {Fritschi}, \citenamefont {Kaiser}, \citenamefont {Klix},
  \citenamefont {Maret},\ and\ \citenamefont {Keim}}]{Keim17}%
  \BibitemOpen
  \bibfield  {author} {\bibinfo {author} {\bibfnamefont {B.}~\bibnamefont
  {Illing}}, \bibinfo {author} {\bibfnamefont {S.}~\bibnamefont {Fritschi}},
  \bibinfo {author} {\bibfnamefont {H.}~\bibnamefont {Kaiser}}, \bibinfo
  {author} {\bibfnamefont {C.~L.}\ \bibnamefont {Klix}}, \bibinfo {author}
  {\bibfnamefont {G.}~\bibnamefont {Maret}},\ and\ \bibinfo {author}
  {\bibfnamefont {P.}~\bibnamefont {Keim}},\ }\bibfield  {title} {\bibinfo
  {title} {Mermin--wagner fluctuations in 2d amorphous solids},\ }\href@noop {}
  {\bibfield  {journal} {\bibinfo  {journal} {Proceedings of the National
  Academy of Sciences}\ }\textbf {\bibinfo {volume} {114}},\ \bibinfo {pages}
  {1856} (\bibinfo {year} {2017})}\BibitemShut {NoStop}%
\bibitem [{\citenamefont {Desmond}\ and\ \citenamefont
  {Weeks}(2009)}]{Weeks09}%
  \BibitemOpen
  \bibfield  {author} {\bibinfo {author} {\bibfnamefont {K.~W.}\ \bibnamefont
  {Desmond}}\ and\ \bibinfo {author} {\bibfnamefont {E.~R.}\ \bibnamefont
  {Weeks}},\ }\bibfield  {title} {\bibinfo {title} {Random close packing of
  disks and spheres in confined geometries},\ }\href
  {https://doi.org/10.1103/PhysRevE.80.051305} {\bibfield  {journal} {\bibinfo
  {journal} {Phys. Rev. E}\ }\textbf {\bibinfo {volume} {80}},\ \bibinfo
  {pages} {051305} (\bibinfo {year} {2009})}\BibitemShut {NoStop}%
\bibitem [{\citenamefont {Nelson}\ and\ \citenamefont
  {Halperin}(1979)}]{Halperin79}%
  \BibitemOpen
  \bibfield  {author} {\bibinfo {author} {\bibfnamefont {D.~R.}\ \bibnamefont
  {Nelson}}\ and\ \bibinfo {author} {\bibfnamefont {B.}~\bibnamefont
  {Halperin}},\ }\bibfield  {title} {\bibinfo {title} {Dislocation-mediated
  melting in two dimensions},\ }\href@noop {} {\bibfield  {journal} {\bibinfo
  {journal} {Physical Review B}\ }\textbf {\bibinfo {volume} {19}},\ \bibinfo
  {pages} {2457} (\bibinfo {year} {1979})}\BibitemShut {NoStop}%
\bibitem [{\citenamefont {Ediger}(2000)}]{Ediger00}%
  \BibitemOpen
  \bibfield  {author} {\bibinfo {author} {\bibfnamefont {M.~D.}\ \bibnamefont
  {Ediger}},\ }\bibfield  {title} {\bibinfo {title} {Spatially heterogeneous
  dynamics in supercooled liquids},\ }\href@noop {} {\bibfield  {journal}
  {\bibinfo  {journal} {Annual review of physical chemistry}\ }\textbf
  {\bibinfo {volume} {51}},\ \bibinfo {pages} {99} (\bibinfo {year}
  {2000})}\BibitemShut {NoStop}%
\bibitem [{\citenamefont {Dasgupta}\ \emph {et~al.}(1991)\citenamefont
  {Dasgupta}, \citenamefont {Indrani}, \citenamefont {Ramaswamy},\ and\
  \citenamefont {Phani}}]{Dasgupta91}%
  \BibitemOpen
  \bibfield  {author} {\bibinfo {author} {\bibfnamefont {C.}~\bibnamefont
  {Dasgupta}}, \bibinfo {author} {\bibfnamefont {A.}~\bibnamefont {Indrani}},
  \bibinfo {author} {\bibfnamefont {S.}~\bibnamefont {Ramaswamy}},\ and\
  \bibinfo {author} {\bibfnamefont {M.}~\bibnamefont {Phani}},\ }\bibfield
  {title} {\bibinfo {title} {Is there a growing correlation length near the
  glass transition?},\ }\href@noop {} {\bibfield  {journal} {\bibinfo
  {journal} {Europhysics Letters}\ }\textbf {\bibinfo {volume} {15}},\ \bibinfo
  {pages} {307} (\bibinfo {year} {1991})}\BibitemShut {NoStop}%
\bibitem [{\citenamefont {Berthier}\ \emph {et~al.}(2011)\citenamefont
  {Berthier}, \citenamefont {Biroli}, \citenamefont {Bouchaud},\ and\
  \citenamefont {Jack}}]{Jack11}%
  \BibitemOpen
  \bibfield  {author} {\bibinfo {author} {\bibfnamefont {L.}~\bibnamefont
  {Berthier}}, \bibinfo {author} {\bibfnamefont {G.}~\bibnamefont {Biroli}},
  \bibinfo {author} {\bibfnamefont {J.-P.}\ \bibnamefont {Bouchaud}},\ and\
  \bibinfo {author} {\bibfnamefont {R.~L.}\ \bibnamefont {Jack}},\ }\href@noop
  {} {\emph {\bibinfo {title} {Overview of different characterisations of
  dynamic heterogeneity}}},\ Vol.\ \bibinfo {volume} {150}\ (\bibinfo
  {publisher} {Oxford University Press Oxford},\ \bibinfo {year} {2011})\
  p.~\bibinfo {pages} {68}\BibitemShut {NoStop}%
\bibitem [{\citenamefont {Paul}\ \emph {et~al.}(2023)\citenamefont {Paul},
  \citenamefont {Mutneja}, \citenamefont {Nandi},\ and\ \citenamefont
  {Karmakar}}]{Karmakar23}%
  \BibitemOpen
  \bibfield  {author} {\bibinfo {author} {\bibfnamefont {K.}~\bibnamefont
  {Paul}}, \bibinfo {author} {\bibfnamefont {A.}~\bibnamefont {Mutneja}},
  \bibinfo {author} {\bibfnamefont {S.~K.}\ \bibnamefont {Nandi}},\ and\
  \bibinfo {author} {\bibfnamefont {S.}~\bibnamefont {Karmakar}},\ }\bibfield
  {title} {\bibinfo {title} {Dynamical heterogeneity in active glasses is
  inherently different from its equilibrium behavior},\ }\href@noop {}
  {\bibfield  {journal} {\bibinfo  {journal} {Proceedings of the National
  Academy of Sciences}\ }\textbf {\bibinfo {volume} {120}},\ \bibinfo {pages}
  {e2217073120} (\bibinfo {year} {2023})}\BibitemShut {NoStop}%
\bibitem [{\citenamefont {Dey}\ \emph {et~al.}(2024)\citenamefont {Dey},
  \citenamefont {Bhattacharya},\ and\ \citenamefont {Karmakar}}]{Karmakar24}%
  \BibitemOpen
  \bibfield  {author} {\bibinfo {author} {\bibfnamefont {S.}~\bibnamefont
  {Dey}}, \bibinfo {author} {\bibfnamefont {A.}~\bibnamefont {Bhattacharya}},\
  and\ \bibinfo {author} {\bibfnamefont {S.}~\bibnamefont {Karmakar}},\
  }\bibfield  {title} {\bibinfo {title} {Enhanced long wavelength mermin-wagner
  fluctuations in two-dimensional active crystals and glasses},\ }\href@noop {}
  {\bibfield  {journal} {\bibinfo  {journal} {arXiv preprint arXiv:2402.10625}\
  } (\bibinfo {year} {2024})}\BibitemShut {NoStop}%
\bibitem [{\citenamefont {Nandi}\ \emph {et~al.}(2018)\citenamefont {Nandi},
  \citenamefont {Mandal}, \citenamefont {Bhuyan}, \citenamefont {Dasgupta},
  \citenamefont {Rao},\ and\ \citenamefont {Gov}}]{Nandi18}%
  \BibitemOpen
  \bibfield  {author} {\bibinfo {author} {\bibfnamefont {S.~K.}\ \bibnamefont
  {Nandi}}, \bibinfo {author} {\bibfnamefont {R.}~\bibnamefont {Mandal}},
  \bibinfo {author} {\bibfnamefont {P.~J.}\ \bibnamefont {Bhuyan}}, \bibinfo
  {author} {\bibfnamefont {C.}~\bibnamefont {Dasgupta}}, \bibinfo {author}
  {\bibfnamefont {M.}~\bibnamefont {Rao}},\ and\ \bibinfo {author}
  {\bibfnamefont {N.~S.}\ \bibnamefont {Gov}},\ }\bibfield  {title} {\bibinfo
  {title} {A random first-order transition theory for an active glass},\
  }\href@noop {} {\bibfield  {journal} {\bibinfo  {journal} {Proceedings of the
  National Academy of Sciences}\ }\textbf {\bibinfo {volume} {115}},\ \bibinfo
  {pages} {7688} (\bibinfo {year} {2018})}\BibitemShut {NoStop}%
\end{thebibliography}
%

\end{document}